%% file: main.tex
\documentclass{article} %
\usepackage{iclr2025_conference,times}
\iclrfinalcopy
\input{math_commands.tex}

\usepackage{hyperref}
\usepackage{url}

\definecolor{zncolor}{rgb}{0.7,0.1,0.4}
\definecolor{gzcolor}{RGB}{255, 0, 255}
\definecolor{tbkcolor}{RGB}{0,100,50}
\definecolor{jmcolor}{rgb}{0.7,0.3,0.7}
\definecolor{njbcolor}{rgb}{0.1,0.4,0.8}

\definecolor{dmd}{HTML}{FBBD00}
\definecolor{gen}{HTML}{C00000}
\definecolor{gan}{HTML}{92D050}
\definecolor{dsm}{HTML}{2E5F7F}
\definecolor{gout}{HTML}{2E5F7F}
\newcommand{\eref}[1]{(\ref{#1})}
\newcommand{\aref}[1]{Appendix~\ref{#1}}
\newcommand{\sref}[1]{Section~\ref{#1}}
\newcommand{\fref}[1]{Fig.~\ref{#1}}
\newcommand{\tref}[1]{Table~\ref{#1}}
\usepackage{cite}
\usepackage{amsmath,amssymb}
\usepackage{amsfonts}
\usepackage{graphicx}
\usepackage{wrapfig}
\usepackage{array}
\newcolumntype{P}[1]{>{\centering\arraybackslash}p{#1}}
\usepackage{algorithm}
\usepackage{algorithmic}
\usepackage{textcomp}
\usepackage{graphicx, booktabs}
\usepackage{listings}
\usepackage{xcolor}
\usepackage[inline,shortlabels]{enumitem}
\usepackage{multirow}
\usepackage{calc}
\usepackage{hyperref}
\usepackage{tcolorbox}
\usepackage{enumitem}
\usepackage{textcomp} 
\usepackage{pythonhighlight}

\definecolor{codegreen}{rgb}{0,0.6,0}
\definecolor{codegray}{rgb}{0.5,0.5,0.5}
\definecolor{codepurple}{rgb}{0.58,0,0.82}
\definecolor{backcolour}{rgb}{1,1,1}

\lstdefinestyle{mystyle}{
    backgroundcolor=\color{backcolour},   
    commentstyle=\color{codegreen},
    keywordstyle=\color{magenta},
    numberstyle=\tiny\color{codegray},
    stringstyle=\color{codepurple},
    basicstyle=\ttfamily\footnotesize,
    breakatwhitespace=false,         
    breaklines=true,                 
    captionpos=b,                    
    keepspaces=true,                 
    numbers=left,                    
    numbersep=6pt,                  
    showspaces=false,                
    showstringspaces=false,
    showtabs=false,                  
    tabsize=1
}

\newcommand\rebuttal[1]{\textcolor{black}{#1}}

\title{\textbf{\textit{Presto!}} Distilling Steps and Layers \\ for Accelerating Music Generation}

\author{Zachary Novack\thanks{ Work done while an intern at Adobe. Correspondence to \texttt{znovack@ucsd.edu} and \texttt{njb@ieee.org}. } \\
UC -- San Diego\\
\And
Ge Zhu \& Jonah Casebeer\\
Adobe Research \\
\AND
Julian McAuley \& Taylor Berg-Kirkpatrick \\
UC -- San Diego\\
\And
Nicholas J. Bryan \\
Adobe Research \\
}

\begin{document}

\maketitle

\begin{abstract}
Despite advances in diffusion-based text-to-music (TTM) methods, efficient, high-quality generation remains a challenge. 
We introduce \textbf{\textit{Presto!}}, an approach to inference acceleration for score-based diffusion transformers via reducing both sampling steps and cost per step. 
To reduce steps, we develop a new score-based distribution matching distillation (DMD) method for the EDM-family of diffusion models, the first GAN-based distillation method for TTM. 
To reduce the cost per step, we develop a simple, but powerful improvement to a recent layer distillation method 
that improves learning 
via better preserving hidden state variance. 
Finally, we combine our step and layer distillation methods together for a dual-faceted approach.
We evaluate our step and layer distillation methods independently and show each yield best-in-class performance. 
Our combined distillation method can generate high-quality outputs with improved diversity, accelerating our base model by 10-18x (230/435ms latency for 32 second mono/stereo 44.1kHz, 15x faster than the comparable SOTA model)
--- the fastest TTM to our knowledge. 
\end{abstract}

\section{Introduction}

We have seen a renaissance of audio-domain generative media \citep{chen2023musicldm, agostinelli2023musiclm, liu2023audioldm, copet2023simple}, with increasing capabilities for both Text-to-Audio (TTA) and Text-to-Music (TTM) generation. 
This work has been driven in-part by audio-domain \emph{diffusion models} \citep{song2020denoising, ho2020denoising, Song2020ScoreBasedGM}, enabling considerably better audio modeling than generative adversarial network (GAN) or variational autoencoder (VAE) methods~\citep{dhariwal2021diffusion}.  
Diffusion models, however, suffer from long inference times due to their iterative denoising process, requiring a substantial number of function evaluations (NFE) 
during inference (i.e.~sampling) and resulting in $\approx$5-20 seconds at best for non-batched \rebuttal{$\approx$32s outputs}. 

Accelerating diffusion inference typically focuses on \emph{step distillation}, 
i.e.~the process of reducing the \emph{number} of sampling steps by distilling the diffusion model into a few-step generator. Methods include consistency-based ~\citep{salimans2022progressive, song2023consistency, Kim2023ConsistencyTM} and adversarial~\citep{Sauer2023AdversarialDD, yin2023one, yin2024improved, kang2024distilling} approaches.
Others have also investigated \emph{layer-distillation}~\citep{Ma2024LearningtoCacheAD, wimbauer2024cache, Moon2024ASE}, which draws from transformer early exiting~\citep{hou2020dynabert, schuster2021consistent} by dropping interior layers to reduce the \emph{cost} per sampling step for image generation.
For TTA/TTM models, however, distillation techniques have only been applied to shorter or lower-quality audio~\citep{bai2023accelerating, Novack2024DITTO2DD}, 
necessitate
$\approx$10 steps (vs. 1-4 step image methods)
to match base quality~\citep{Saito2024SoundCTMUS}, and have not successfully used layer or GAN-based distillation methods.

\begin{figure}
    \centering
    \includegraphics[width=0.9\linewidth]{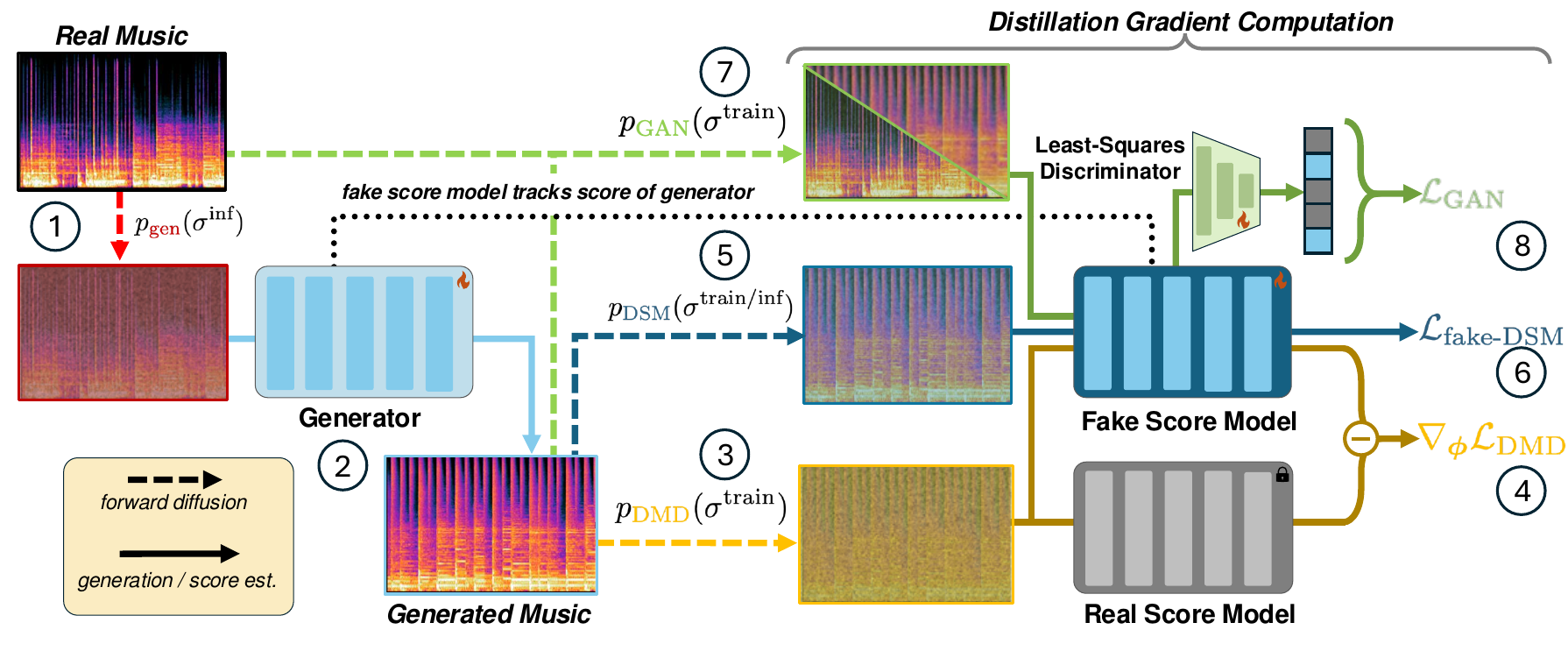}
    \vspace{-0.1cm}
    \caption{\textbf{Presto-S}. Our goal is to distill the initial ``real'' score model (grey) $\mu_{\boldsymbol{\theta}}$ into a few-step generator (light blue) $G_{\boldsymbol\phi}$ to minimize the KL divergence between the distribution of $G_{\boldsymbol\phi}$'s outputs and the real distribution. This requires that we train an auxillary ``fake'' score model $\mu_{\boldsymbol{\psi}}$ (dark blue) that estimates the score of the \emph{generator's} distribution at each gradient step. Formally: (1) real audio is corrupted with Gaussian noise sampled from the generator noise distribution $p_{\text{\textcolor{gen}{gen}}}(\sigma^{\text{inf}})$ which is then (2) passed into the generator to get its output. Noise is then added to this generation according to three \emph{different} noise distributions: (3)  $p_{\text{\textcolor{dmd}{DMD}}}(\sigma^{\text{train}})$, which is (4) passed into both the real and fake score models  to calculate the distribution matching gradient $\nabla_\phi\mathcal{L}_{\text{DMD}}$; (5) $p_{\text{\textcolor{dsm}{DSM}}}(\sigma^{\text{train/inf}})$, which is used to (6) train the fake score model on the \emph{generator's} distribution with $\mathcal{L}_{\text{fake-DSM}}$; and (7) an adversarial distribution $p_{\text{\textcolor{gan}{GAN}}}(\sigma^{\text{train}})$, which along with the real audio is (8) passed into a least-squares discriminator built on the fake score model's intermediate activations to calculate $\mathcal{L}_{\text{GAN}}$.}
    \label{fig:presto_s}
\end{figure}

We present \textbf{Presto}\footnote{\emph{Presto} is the common musical term denoting fast music from 168-200 beats per minute.}, a dual-faceted distillation approach to inference acceleration for score-based diffusion transformers via reducing the number of sampling steps and the cost per step. 
\textbf{Presto} includes three distillation methods:
(1) \textbf{Presto-S}, a new distribution matching distillation algorithm for \emph{score-based}, EDM-style diffusion models
(see ~\fref{fig:presto_s}) leveraging GAN-based step distillation with the flexibility of \emph{continuous-time} models,
(2) \textbf{Presto-L}, a conditional layer distillation method designed to better preserve hidden state variance during distillation, and 
(3) \textbf{Presto-LS}, a combined layer-step distillation method that critically uses layer distillation and \emph{then} step distillation while disentangling layer distillation from  real and fake score-based gradient estimation.

To evaluate our approach, we ablate the design space for both distillation processes. First, we show our step distillation method achieves best-in-class acceleration and quality via careful choice of loss noise distributions, GAN design, and continuous-valued inputs, the first such method to match base TTM diffusion sampling quality with 4-step inference. 
Second, we show our layer distillation method offers a consistent improvement in both speed \emph{and} performance over SOTA layer dropping methods and base diffusion sampling.
Finally, we show that layer-step distillation accelerates our base model by 10-18x (230/435ms latency for 32 second mono/stereo 44.1kHz, 15x faster than the comparable SOTA model) while notably improving diversity over step-only distillation.

Overall, our core contributions include the development of a holistic approach to accelerating score-based diffusion transformers including
: \textbf{(1)} The development of distribution matching distillation for continuous-time score-based diffusion (i.e. EDM), the first GAN-based distillation method for TTM. \textbf{(2)} The development of an improved layer distillation method that consistently improves upon both past layer distillation method and our base diffusion model. \textbf{(3)} The development of the first combined layer and step distillation method.
    \textbf{(4)} Evaluation showing our step, layer, and layer-step distillation methods are all best-in-class and, when combined, can accelerate our base model by 10-18x (230/435ms latency for 32 second mono/stereo 44.1kHz, 15x faster than Stable Audio Open~\citep{evans2024open}), the fastest TTM model to our knowledge. For sound examples (anonymous link), see \url{https://presto-music.github.io/web/}.

\section{Background \& Related Work}

\subsection{Music Generation}
Audio-domain music generation methods commonly use autoregressive (AR) techniques~\citep{zeghidour2021soundstream, agostinelli2023musiclm, copet2023simple} or diffusion~\citep{forsgren2022riffusion, liu2023audioldm, liu2023audioldm2, schneider2023mo}. Diffusion-based TTA/TTM~\citep{forsgren2022riffusion, liu2023audioldm, liu2023audioldm2, schneider2023mo, stableaudio} has shown the promise of full-text control~\citep{huang2023noise2music}, precise musical attribute control~\citep{ Novack2024Ditto, Novack2024DITTO2DD, Tal2024JointAA}, structured long-form generation~\citep{Evans2024LongformMG}, and higher overall quality over AR methods~\citep{stableaudio, Evans2024LongformMG, Novack2024Ditto, evans2024open}. The main downside of diffusion, however, is that it is slow and thus not amenable to interactive-rate control.

\subsection{Score-Based Diffusion Models}
Continuous-time diffusion models have shown great promise over discrete-time models both for their improved performance on images~\citep{balaji2022ediffi, Karras2023AnalyzingAI, liu2024playgroundv3} \emph{and} audio~\citep{nistal2024diff, zhu2023edmsound, Saito2024SoundCTMUS}, as well as their relationship to the general class of flow-based models~\citep{Sauer2024FastHI, Tal2024JointAA}.
Such models involve a forward noising process that gradually adds Gaussian noise to real audio signals $\bm{x}_{\text{real}}$ and 
a reverse process that transforms pure Gaussian noise back into data~\citep{Song2020ScoreBasedGM, sohl2015deep}. The reverse process is defined by a stochastic differential equation (SDE) with an equivalent ordinary differential equation (ODE) form called the \emph{probability flow} (PF) ODE \citep{Song2020ScoreBasedGM}:
\begin{equation}
    \mathrm{d}\bm{x} = -\sigma \nabla_{\bm{x}} \log p(\bm{x} \mid \sigma )\mathrm{d}\sigma,
\end{equation}
where $\nabla_{\bm{x}} \log p(\bm{x} \mid \sigma )$ is the score function of the marginal density of $\bm{x}$ (i.e.~the noisy data) at noise level $\sigma$ according to the forward diffusion process. 
Thus, the goal of score-based diffusion models is to learn a \emph{denoiser} network $\mu_{\boldsymbol{\theta}}$ such that $\mu_{\boldsymbol{\theta} }(\bm{x}, \sigma) = \mathbb{E}[\bm{x}_{\text{real}} \mid \bm{x}, \sigma]$. The score function is:
\begin{equation}
    \nabla_{\bm{x}} \log p(\bm{x} \mid \sigma ) \approx \frac{\bm{x} - \mu_{\boldsymbol{\theta}} (\bm{x},\sigma)}{\sigma}.
\end{equation}
Given a trained score model, we can generate samples at inference time by setting a decreasing \emph{noise schedule} of $N$ levels $\sigma_{\text{max}} = \sigma_{N}> \sigma_{N-1} > \dots >  \sigma_{0} = \sigma_{\text{min}}$ and iteratively solving the ODE at these levels using our model and any off-the-shelf ODE solver (e.g.~Euler, Heun). 

The EDM-family~\citep{karras2022elucidating, Karras2023AnalyzingAI} of score-based diffusion models is of particular interest and unifies several continuous-time model variants within a common framework and improves model parameterization and training process. 
The EDM score model is trained by minimizing a reweighted denoising score matching (DSM) loss \citep{Song2020ScoreBasedGM}:
\begin{equation}
    \mathcal{L}_{\text{DSM}} = \mathbb{E}_{\bm{x}_{\text{real}} \sim \mathcal{D}, \sigma \sim p(\sigma^{\text{train}}), \epsilon \sim \mathcal{N}(0, \boldsymbol{I})}\left[\lambda(\sigma)\|\bm{x}_{\text{real}} - \mu_{\boldsymbol{\theta}}(\bm{x}_{\text{real}} + \epsilon\sigma, \sigma)\|_2^2\right],
\end{equation}
where $p(\sigma^{\text{train}})$ denotes the \emph{noise distribution} during training,  and $\lambda(\sigma)$ is a noise-level weighting function. Notably, EDM defines a \emph{different} noise distribution to discretize for inference $p(\sigma^{\text{inf}})$ that is distinct from $p(\sigma^{\text{train}})$ (see~\fref{fig:signoise}), as opposed to a noise schedule shared between training and inference.
Additionally, EDMs represent the denoising network using extra noise-dependent preconditioning parameters, training a network $f_{\boldsymbol{\theta}}$ with the parameterization:
\begin{equation}
        \mu_{\boldsymbol{\theta}}(\bm{x}, \sigma) = c_{\text{skip}}(\sigma) \bm{x} + c_{\text{out}}(\sigma)f_{\boldsymbol{\theta}}( c_{\text{in}}(\sigma) \bm{x}, c_{\text{noise}}(\sigma)).
\end{equation}
For TTM models, $\mu_{\boldsymbol{\theta} }$ is equipped with various condition embeddings (e.g.~text) $\mu_{\boldsymbol{\theta} }(\bm{x}, \sigma, \bm{e})$. To increase text relevance and quality at the cost of diversity, we employ \emph{classifier free guidance} (CFG) \citep{ho2022classifier}, converting the denoised output to: $\tilde{\mu}^w_{\boldsymbol{\theta} }(\bm{x}, \sigma, \bm{e}) = \mu_{\boldsymbol{\theta} }(\bm{x}, \sigma, \boldsymbol{\emptyset}) + w (\mu_{\boldsymbol{\theta} }(\bm{x}, \sigma, \bm{e}) - \mu_{\boldsymbol{\theta} }(\bm{x}, \sigma, \boldsymbol{\emptyset}))$, where $w$ is the guidance weight and $\boldsymbol{\emptyset}$ is a ``null'' conditioning.

\subsection{Diffusion Distillation}
Step distillation is the process of reducing diffusion sampling steps by distilling a base model into a few-step generator. Such methods can be organized into two broad categories. 
Online consistency approaches such as consistency models~\citep{song2023consistency}, consistency trajectory models~\citep{Kim2023ConsistencyTM}, and variants~\citep{Ren2024HyperSDTS, Wang2024PhasedCM} distill directly by enforcing consistency across the diffusion trajectory and optionally include an adversarial loss~\citep{Kim2023ConsistencyTM}.
While such approaches have strong 1-step generation for images, attempts for audio have been less successful and only capable of generating short segment (i.e.~$<10$ seconds), applied to lower-quality base models limiting upper-bound performance, needing up to 16 sampling steps to match baseline quality (still slow), 
and/or did not successfully leverage adversarial losses  which have been found to increase realism for other domains~\citep{bai2023accelerating, Saito2024SoundCTMUS, Novack2024DITTO2DD}. 

In contrast, offline adversarial distillation methods include Diffusion2GAN~\citep{kang2024distilling}, LADD~\citep{Sauer2024FastHI}, and DMD~\citep{yin2023one}. 
Such methods work by generating large amounts of offline noise--sample pairs from the base model, and finetuning the model into a conditional GAN for few-step synthesis.
These methods can surpass their adversarial-free counterparts, yet require expensive offline data generation and massive compute infrastructure to be efficient. 

Alternatively, improved DMD (DMD2)~\citep{yin2024improved} introduces an online adversarial diffusion distillation method for images. 
DMD2 (1) removes the need for expensive offline data generation (2) adds a GAN loss and (3) outperforms consistency-based methods and improves overall quality. 
DMD2 primarily works by distilling a one- or few-step generator $G_{\boldsymbol{\phi}}$ from a base diffusion model $\mu_{\text{real}}$, while simultaneously learning a score model of the generator's distribution online $\mu_{\text{fake}}$ in order to approximate a target KL objective (with $\mu_{\text{real}}$) used to train the generator.  
To our knowledge, there are no adversarial diffusion distillation methods for TTM or TTA.

Beyond step distillation, layer distillation, or the process of dropping interior layers to reduce the cost per sampling step, has been recently studied~\citep{Moon2024ASE, wimbauer2024cache}. Layer distillation  draws inspiration from transformer early exiting and layer caching~\citep{hou2020dynabert, schuster2021consistent} and has found success for image diffusion, but has not been compared or combined with step distillation methods and has not been developed for TTA/TTM. In our work, we seek to understand how step and layer distillation interact for accelerating music generation. 

\section{\textbf{\textit{Presto!}}}
We propose a dual-faceted distillation approach for inference acceleration of continuous-time diffusion models. 
Continuous-time models have been shown to outperform discrete-time DDPM models~\citep{ song2020denoising, karras2022elucidating, karras2024analyzing}, but past DMD/DMD2 work focuses on the latter.
Thus, we redefine DMD2 (a step distillation method) in~\sref{sec:edmdmd}  for continuous-time score models, then present
an improved formulation and study its design space in~\sref{sec:presto-s}. 
Second, we design a simple, but powerful improvement to the SOTA layer distillation method to understand the impact of reducing inference cost per step in~\sref{sec:codda}. 
Finally, we investigate how to combine step and layer distillation methods together in~\sref{sec:lsdistill}.

\subsection{EDM-Style Distribution Matching Distillation}\label{sec:edmdmd}
We first redefine DMD2 in the language of continuous-time, score-based diffusion models (i.e. EDM-style).
Our goal is to 
distill
our score model $\mu_{\boldsymbol{\theta}}$ (which we equivalently denote as $\mu_{\text{real}}$, as it is trained to model the score of real data) into an accelerated
generator $G_{\boldsymbol{\phi}}$ 
that can sample in 1-4 steps.  
Formally, we wish to minimize the reverse KL Divergence between the real distribution $p_{\text{real}}$ and the generator $G_{\boldsymbol{\phi}}$'s distribution $p_{\text{fake}}$: $\mathcal{L}_{\text{DMD}} = D(p_{\text{fake}} \| p_{\text{real}})$.
The KL term cannot be calculated explicitly, but we can calculate its \emph{gradient} with respect to the generator if we can access the score of the generator's distribution.
Thus, we  also train a ``fake" score model $\mu_{\boldsymbol{\psi}}$ (or equivalently, $\mu_{\text{fake}}$) to approximate the generator distribution's \emph{score function} at each gradient step during training.

First, given some real data $\bm{x}_{\text{real}}$, we sample a noise level from a set of predefined levels $\sigma \sim \{\sigma_i\}_{\text{gen}}$, 
and then pass the corrupted real data through the generator to get the generated output $\hat{\bm{x}}_{\text{gen}} = G_{\boldsymbol{\phi}}(\bm{x}_{\text{real}} + \sigma\epsilon, \sigma)$, where $\epsilon \sim \mathcal{N}(0, \boldsymbol{I})$ (we omit the conditioning $\bm{e}$ for brevity). 
The gradient of the KL divergence between the real and the generator's distribution can then be calculated as:
\begin{equation}\label{eq:dmd}
    \nabla_{\boldsymbol{\phi}} \mathcal{L}_{\text{DMD}} =\mathbb{E}_{\substack{\sigma \sim \{\sigma_i\}, \epsilon \sim \mathcal{N}(0, \boldsymbol{I})}} \left[\left((\mu_{\text{fake}}(\hat{\bm{x}}_{\text{gen}} + \sigma\epsilon, \sigma) - \tilde\mu_{\text{real}}^w(\hat{\bm{x}}_{\text{gen}} + \sigma\epsilon, \sigma )\right)\nabla_{\boldsymbol{\phi}}\hat{\bm{x}}_{\text{gen}}\right],
\end{equation} 
where $\{\sigma_i\}$ are the predefined noise levels for all loss calculations, and $\tilde\mu_{\text{real}}^w$ is the \emph{CFG-augmented} real score model. To ensure that $\mu_{\text{fake}}$ accurately models the score of the generator's distribution at each gradient update, we train the fake score model with the weighted-DSM loss (i.e.~standard diffusion training), but on \emph{the generator outputs}:
\begin{equation}\label{eq:dsm}
    \arg \min_{\boldsymbol{\psi}} \mathcal{L}_{\text{fake-DSM}} = \mathbb{E}_{\sigma \sim \{\sigma_i\}, \epsilon \sim \mathcal{N}(0, \boldsymbol{I})}\left[\lambda(\sigma)\|\hat{\bm{x}}_{\text{gen}} - \mu_{\text{fake}}(\hat{\bm{x}}_{\text{gen}} + \sigma\epsilon, \sigma)\|_2^2\right]
\end{equation}
To avoid using offline data \citep{yin2023one}, the fake score model is updated \emph{5 times as often} as the generator to stabilize the estimation of the generator's distribution. 
DMD2 additionally includes an explicit adversarial loss in order to improve quality. Specifically, a discriminator head $D_{\boldsymbol{\psi}}$ is attached to the intermediate feature activations
of the fake score network $\mu_{\text{fake}}$, and thus is trained with the nonsaturating GAN loss:
\begin{equation}\label{eq:gan}
    \arg 
    \min_{\boldsymbol{\phi}} \max_{\boldsymbol{\psi}} \mathbb{E}_{\substack{\sigma \sim \{\sigma_i\}, \\ \epsilon \sim \mathcal{N}(0, \boldsymbol{I})}}[\log D_{\boldsymbol{\psi}}(\bm{x}_{\text{real}} + \sigma\epsilon, \sigma)] + \mathbb{E}_{\substack{\sigma \sim \{\sigma_i\}, \\ \epsilon \sim \mathcal{N}(0, \boldsymbol{I})}}[-\log D_{\boldsymbol{\psi}}(\hat{\bm{x}}_{\text{gen}} + \sigma\epsilon, \sigma)],
\end{equation}
which follows past work on using diffusion model backbones as discriminators \citep{Sauer2024FastHI}. In all, the generator $G_{\boldsymbol{\phi}}$ is thus trained with a combination of the distribution matching loss $\mathcal{L}_{\text{DMD}}$ and the adversarial loss $\mathcal{L}_{\text{GAN}}$, while the fake score model (and its discriminator head) is trained with the fake DSM loss $\mathcal{L}_{\text{fake-DSM}}$ and the adversarial loss $\mathcal{L}_{\text{GAN}}$. To sample from the distilled generator, DMD2 uses consistency model-style ``ping-pong sampling" \citep{song2023consistency}, where the model iteratively denoises (starting at pure noise $\sigma_{\text{max}}$) and \emph{renoises} to progressively smaller noise levels. 

Regarding past work, we note~\citet{yin2024improved} \emph{did} present a small-scale EDM-style experiment, 
but treated EDM models as if they were functions of discrete noise timesteps. 
This re-discretization runs counterintuitive to using score-based models for distribution matching, since the fake and real score models are meant to be run and trained in continuous-time and can adapt to variable points along the noise process. 
Furthermore, this disregards the ability of continuous-time models to capture the \emph{entire} noise process from noise to data and enable \emph{exact} likelihoods rather than ELBOs \citep{Song2020ScoreBasedGM}. 
Additionally, since DMD2 implicitly aims to learn an integrator of the PF ODE 
$G_{\boldsymbol{\phi}}(\bm{x}, \sigma) \approx \bm{x} + \int_{\sigma}^{\sigma_{\text{min}}} -\delta\nabla \log p(\bm{x} \mid \delta) \textrm{d}\delta$ (like other data-prediction distillation methods \citep{song2023consistency}), learning this integral for any small set $\{\sigma_i\}$ restricts the generator's modeling capacity.

\subsection{Presto-S: Score-based Distribution Matching Distillation}
\label{sec:presto-s}
We develop our \emph{score-based} distribution matching step distillation, \textbf{Presto-S} below and in~\fref{fig:presto_s} as well as the algorithm in~\aref{app:presto_s_algo}, \rebuttal{a pseudo-code walkthrough in~\aref{app:walkthrough}, and expanded visualization in~\aref{app:expanded_diagram}}.

\subsubsection{Continuous-Time Generator Inputs}
In~\sref{sec:edmdmd}, the noise level and/or timestep is sampled from a \emph{discrete}, hand-chosen set $\{\sigma_i\}_{\text{gen}}$.
Discretizing inputs, however, forces the model to 1) be a function of a specific \emph{number} of steps, requiring users to retrain separate models for each desired step budget \citep{yin2024improved,kohler2024imaginefa} and 2) be a function of \emph{specific} noise levels, which may not be optimally aligned with where different structural, semantic, and perceptual features arise in the diffusion process~\citep{Si2023FreeUFL, kynkaanniemi2024applying, balaji2022ediffi, sabour2024align}. 
When extending to continuous-time models,  we train the distilled generator $G_{\boldsymbol{\phi}}$ as a function of the continuous noise level sampled from the \emph{distribution} $\sigma \sim p(\sigma)$. 
This allows our generator to both adapt better to variable budgets and to variable noise levels, as the generator can be trained with all noise levels sampled from $p(\sigma)$.

\subsubsection{Perceptual Loss Weighting with Variable Noise Distributions}\label{sec:losses}

\begin{wrapfigure}{r}{0.45\textwidth}
	\vskip -0.24in
	\centering
	\includegraphics[width=\linewidth]{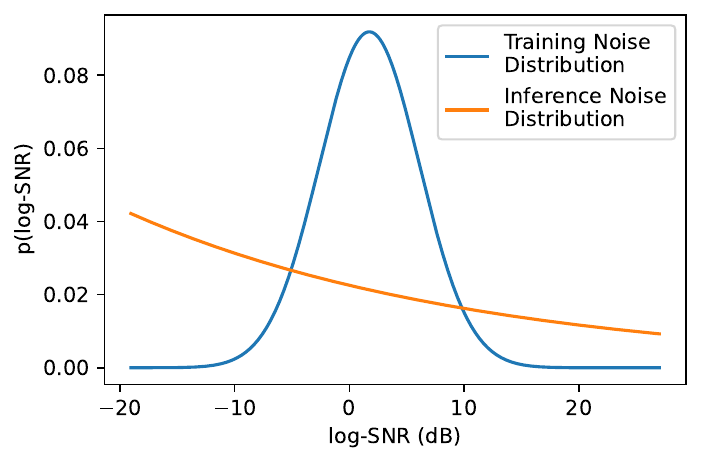}
 \vskip -0.15in
    \caption{Training/Inference distributions for EDM models, in decibel SNR space.}
	\vskip -0.13in
	\label{fig:signoise}
\end{wrapfigure}
A key difference between discrete-time and continuous-time diffusion models is the need for \emph{discretization} of the noise process during inference. 
In discrete models, a single noise schedule defines a particular mapping between timestep $t$ and its noise level $\sigma$, and is fixed throughout training and inference. 
In continuous-time EDM models, however, we use a noise \emph{distribution} $p(\sigma^{\text{train}})$ to sample during training, and a separate noise distribution for inference $p(\sigma^{\text{inf}})$ that is discretized to define the sampling schedule. 
In particular, when viewed in terms of the \emph{signal-to-noise ratio} $1/\sigma^2$ or SNR as shown in~\fref{fig:signoise}, the \emph{training} noise distribution puts the majority of its mass in the mid-to-high SNR range of the diffusion process. This design choice focuses on semantic and perceptual features, while the \emph{inference} noise distribution is more evenly distributed but with a bias towards the low-SNR region, giving a bias to low-frequency features.

However, recall that \emph{every} loss term including~\eref{eq:dmd}, \eref{eq:dsm}, and \eref{eq:gan} requires an additional re-corruption process that must follow a noise distribution, significantly expanding the design space for score-based models. 
Thus, we disentangle these forward diffusion processes and replace the shared discrete noise set with four \emph{separate noise distributions} $p_{\text{\textcolor{gen}{gen}}}$, $p_{\text{\textcolor{dmd}{DMD}}}$, $p_{\text{\textcolor{dsm}{DSM}}}$, and $p_{\text{\textcolor{gan}{GAN}}}$, corresponding to the inputs to the generator and each loss term respectively, with no restrictions on how each weights each noise level 
(rather than forcing a particular noise weighting for all computation). 

Then, if we apply the original DMD2 method naively to the EDM-style of score-models, we get $p_{\text{\textcolor{gen}{gen}}}(\sigma^{\text{inf}}) = p_{\text{\textcolor{dmd}{DMD}}}(\sigma^{\text{inf}}) = p_{\text{\textcolor{dsm}{DSM}}} (\sigma^{\text{inf}}) = p_{\text{\textcolor{gan}{GAN}}}(\sigma^{\text{inf}})$.
This choice of $p_{\text{\textcolor{gen}{gen}}}(\sigma^{\text{inf}})$  reasonably aligns the generator inputs during distillation to the inference process itself, but each loss noise distribution is somewhat misaligned from its role in the distillation process. 
In particular:
\vspace{-.25cm}
\begin{itemize}
    \item $p_{\text{\textcolor{dmd}{DMD}}}$: The distribution matching gradient is the only point that the generator gets a signal from the \emph{CFG-augmented} outputs of the teacher. CFG is critical for text following, but \emph{primarily} within the mid-to-high SNR region of the noise~\citep{kynkaanniemi2024applying}.
    \item $p_{\text{\textcolor{gan}{GAN}}}$: As in most adversarial distillation methods \citep{Sauer2023AdversarialDD, yin2023one}, the adversarial loss's main strength is to increase the perceptual \emph{realism/quality} of the outputs, which arise in the mid-to-high SNR regions, rather than structural elements.
    \item $p_{\text{\textcolor{dsm}{DSM}}}$: The score model training should in theory mimic standard diffusion training, and may benefit from the training distribution's provably faster convergence~\citep{wang2024evaluating} (as the fake score model is updated \emph{online} to track the generator's distribution).
\end{itemize}
\vspace{-.25cm}
Thus, we shift all of the above terms to use the training distribution $p_{\text{\textcolor{dmd}{DMD}}}(\sigma^{\text{train}}),  p_{\text{\textcolor{dsm}{DSM}}} (\sigma^{\text{train}})$ and $p_{\text{\textcolor{gan}{GAN}}}(\sigma^{\text{train}})$ 
to force the distillation process to focus on perceptually relevant noise regions.

\subsubsection{Audio-Aligned Discriminator Design}

The original DMD2 uses a classic non-saturating GAN loss. The discriminator is a series of convolutional blocks downsampling the intermediate features into a \emph{single} probability for real vs. fake. While this approach is standard in image-domain applications, many recent adversarial waveform synthesis works \citep{kumar2023high, Zhu2024MusicHiFiFH} use a \emph{Least-Squares} GAN loss:
\begin{equation}
    \arg 
    \min_{\boldsymbol{\phi}} \max_{\boldsymbol{\psi}}
    \mathbb{E}_{\substack{\sigma \sim p_{\text{\textcolor{gan}{GAN}}}(\sigma^{\text{train}}), \\ \epsilon \sim \mathcal{N}(0, \boldsymbol{I})}}[\|D_{\boldsymbol{\psi}}(\bm{x}_{\text{real}} + \sigma\epsilon, \sigma)\|_2^2] + \mathbb{E}_{\substack{\sigma \sim p_{\text{\textcolor{gan}{GAN}}}(\sigma^{\text{train}}), \\ \epsilon \sim \mathcal{N}(0, \boldsymbol{I})}}[\|1-D_{\boldsymbol{\psi}}(\hat{\bm{x}}_{\text{gen}} + \sigma\epsilon, \sigma)\|_2^2],
\end{equation}
where the outputs of the discriminator $D_{\boldsymbol{\psi}}$ are only \emph{partially} downsampled into a lower-resolution version of the input data (in this case, a latent 1-D tensor). This forces the discriminator to attend to more fine-grained, temporally-aligned features for determining realness, as the loss is averaged across the partially downsampled discriminator outputs. Hence, we use this style of discriminator for \textbf{Presto-S} to both improve and stabilize~\citep{mao2017least} the GAN gradient into our generator.

\subsection{Presto-L: Variance and Budget-Aware Layer Dropping}\label{sec:codda}
\begin{figure}[h]
    \vspace{-0.15in}
    \begin{center}
            \includegraphics[width=\linewidth]{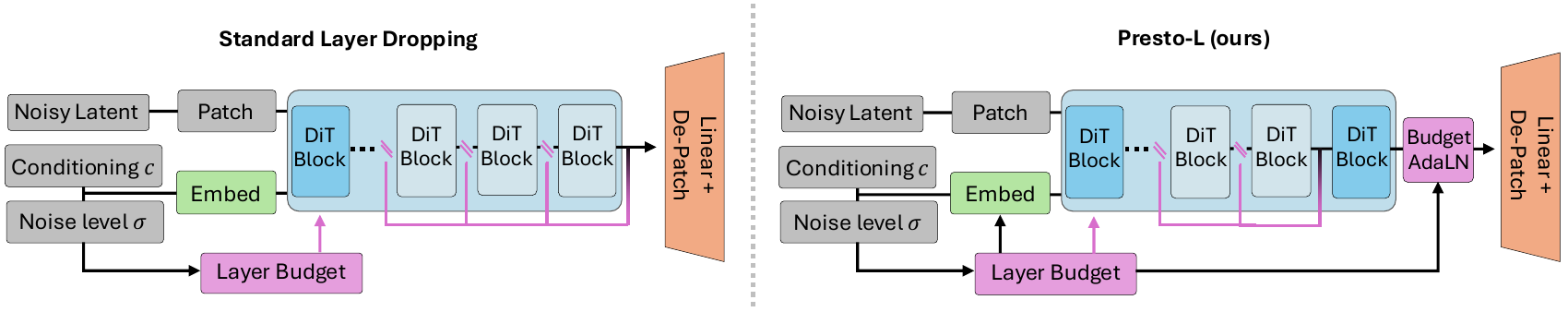}
    \end{center}
    \vspace{-0.4cm}
    \caption{Baseline layer dropping (left) vs. \textbf{Presto-L} (right). Standard layer dropping uses the noise level $\sigma$ to set the budget of layers to drop, starting from the back of the DiT blocks. \textbf{Presto-L} shifts this dropping by one to the second-to-last block and adds explicit budget conditioning.}
    \label{fig:codda}
    \vspace{-0.05in}
\end{figure}

Given our step distillation approach above, we now seek to reduce the cost of individual \emph{steps} themselves through layer distillation, and then combine both step and layer distillation in~\sref{sec:lsdistill}.
We begin with the current SOTA method: ASE~\citep{Moon2024ASE}. ASE employs a fixed dropping schedule that monotonically maps noise levels to compute budgets, allocating more layers to lower noise levels. We enhance this method in three key ways: (1) ensuring consistent variance in layer distilled outputs, (2) implementing explicit budget conditioning, and (3) aligning layer-dropped outputs through direct distillation. \rebuttal{See~\aref{app:presto_l_alg} for more details.}

\begin{wrapfigure}{r}{0.45\textwidth}
	\vskip -0.2in
	\centering
	\includegraphics[width=\linewidth, trim={0cm 0cm 0cm 0cm}, clip]{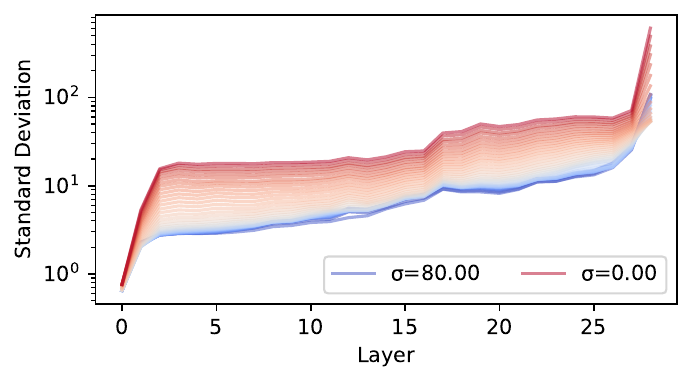}
 \vskip -0.2in
    \caption{Hidden activation variance vs. layer depth. Each line is a unique noise level.}
	\vskip -0.13in
	\label{fig:ditsd}
\end{wrapfigure}

\textbf{Variance Preservation}:
First, we inspect the within-layer activation variance of our base model in~\fref{fig:ditsd}. We find that while the variance predictably
increases over depth, it notably spikes \emph{on the last layer} up to an order of magnitude higher, indicating that the last layer behaves much differently as it is the direct input to the linear de-embedding layer.
ASE, however, always drops layers starting from the \emph{last} layer and working backwards, thus always removing this behavior. 
Hence, we remedy this fact and \emph{shift} the layer dropping schedule by 1 to drop starting at the \emph{second} to last layer, always rerouting back into the final layer to preserve the final layer's behavior.

\textbf{Budget Conditioning}:  
We include \emph{explicit} budget conditioning into the model itself so that the model can directly adapt computation to the budget level.
This conditioning comes in two places: (1) a global budget embedding added to the noise level embedding, thus contributing to the internal Adaptive Layer Norm (AdaLN) conditioning inside the DiT blocks, and (2) an additional AdaLN layer on the outset of the DiT blocks conditional only on the budget, in order to effectively rescale the outputs to account for the change in variance. Following \citep{peebles2023scalable, zhang2023adding}, we zero-initialize both budget conditioning modules to improve finetuning stability.

\textbf{Knowledge Distillation}: To encourage  distillation without holding the base model in memory, we employ a
\emph{self-teacher} loss.
Formally, if $h_{\boldsymbol{L}}(\bm{x}, \boldsymbol{\theta})$ and $h_{\text{full}}(\bm{x}, \boldsymbol{\theta})$ are the normalized outputs of the final DiT layer with and without layer dropping respectively,
the self-teacher loss is $
    \mathcal{L}_{\text{st}} = \|h_{\boldsymbol{L}}(\bm{x}, \boldsymbol{\theta}) - \texttt{sg}(h_{\text{full}}(\bm{x}, \boldsymbol{\theta}))\|_2^2,$
where $\texttt{sg}$ denotes a stop-gradient. This gives additional supervision during the early phases of finetuning so the layer-dropped outputs can match full model performance.

We show the differences between our \textbf{Presto-L} and the baseline approach in Fig.~\ref{fig:codda}. By conditioning directly on the budget, and shifting the dropping schedule to account for the final DiT block behavior, we able to more adapt computation for reduced budgets while preserving performance.

\subsection{Presto-LS: Layer-Step Distillation}\label{sec:lsdistill}

As the act of layer distillation is, in principle, unrelated to the step distillation, there is no reason \emph{a priori} that these methods could not work together.
However, we found combining such methods to be surprisingly non-trivial. 
In particular, we empirically find that attempting both performing \textbf{Presto-L} finetuning and \textbf{Presto-S} at the same time OR performing \textbf{Presto-L} finetuning from an initial \textbf{Presto-S} checkpoint results in large instability and model degradation, as the discriminator dominates the optimization process and achieves near-perfect accuracy on real data. 

We instead find three key factors in making combined step and layer distillation work:
 (1) \emph{Layer-Step Distillation} -- we first perform layer distillation then step distillation, which is more stable as the already-finetuned layer dropping prevents generator collapse; 
 (2) \emph{Full Capacity Score Estimation} -- we keep the real and fake score models initialized from the \emph{original} score model rather than the layer-distilled model, as this stabilizes the distribution matching gradient and provides regularization to the discriminator since the fake score model and the generator are initialized with different weights; and 
 (3) \emph{Reduced Dropping Budget} -- we keep more layers during the layer distillation. We discuss more in~\sref{sec:lsd_res} and how alternatives fail in \aref{app:lsfails}.

\section{Experiments}

We show the efficacy of \textbf{Presto} via a number of experiments. We first ablate the design choices afforded by \textbf{Presto-S}, and separately show how \textbf{Presto-L} flatly improves standard diffusion sampling. We then show how \textbf{Presto-L} and \textbf{Presto-S} stack up against SOTA baselines, and how we can combine such approaches for further acceleration, with both quantitative and subjective metrics. We finish by describing a number of extensions enabled by our accelerated, continuous-time framework.

\subsection{Setup}

\textbf{Model:} We use latent diffusion~\citep{rombach2022high} with a fully convolutional VAE~\citep{kumar2023high} to generate mono 44.1kHz audio and convert to stereo using MusicHiFi~\citep{Zhu2024MusicHiFiFH}. 
Our latent diffusion model builds upon DiT-XL~\citep{peebles2023scalable} and takes in three conditioning signals: the noise level, text prompts, and beat per minute (BPM) for each song. We use FlashAttention-2~\citep{dao2023flashattention} for the DiT and \texttt{torch.compile} for the VAE decoder and MusicHiFi. 
For more details, see \aref{app:arch}.

\textbf{Data:} We use a 3.6K hour dataset of mono 44.1 kHz licensed instrumental music, augmented with pitch-shifting and time-stretching. Data  includes musical meta-data and synthetic captions. 
For evaluation, we use Song Describer (no vocals)~\citep{manco2023song} split into 32 second chunks.

\textbf{Baselines:} We compare against a number of acceleration algorithms \rebuttal{using our base model}: Consistency Models (CM)~\citep{song2023consistency}, SoundCTM~\citep{Saito2024SoundCTMUS}, DITTO-CTM~\citep{Novack2024DITTO2DD}, 
DMD-GAN~\citep{yin2024improved}, and
ASE~\citep{Moon2024ASE}, \rebuttal{as well as}
MusicGen~\citep{copet2023simple}
and Stable Audio Open~\citep{evans2024open}. See \aref{app:bases} for more details.

\textbf{Metrics:} We use a number of common evaluation metrics for text-to-music generation, including distributional quality/diversity metrics (FAD/MMD/Density/Recall/Coverage), prompt adherence (CLAP Score), and latency (RTF). See \aref{app:bases} for more details.

\subsection{Exploring the Design Space of \textbf{Presto-S}}

\begin{wraptable}{r}{8cm}
    \vspace{-0.3cm}
    \centering
    \tiny
    \begin{tabular}{cccc|ccc}
    \toprule
    $p_{\text{\textcolor{gen}{gen}}}$ & $p_{\text{\textcolor{dmd}{DMD}}}$ & $p_{\text{\textcolor{dsm}{DSM}}}$& $p_{\text{\textcolor{gan}{GAN}}}$&  FAD&  MMD& CLAP\\\midrule
 \multicolumn{7}{c}{\textbf{Least-Squares GAN}}\\
    \midrule
Inf.& Inf.& Inf.& Inf.& 0.37& 1.73& 27.45\\
Inf.& Inf.& Tr.& Inf.& 0.37& 1.58& 26.45\\
Inf.& Inf.& Tr.& Tr.& 0.37& 1.51& 24.90\\
Inf.& Tr.& Tr.& Inf.& 0.27& 1.27& 33.12\\
Inf.& Tr.& Inf.& Tr.& \underline{0.23}& \underline{0.86}& \textbf{33.29}\\
Inf.& Tr.& Tr.& Tr.& \textbf{0.22}& \textbf{0.83}& \underline{33.13}\\
Tr.& Tr.& Tr.& Tr.& 0.24& 0.99& 30.89\\
 \midrule
 \multicolumn{7}{c}{\textbf{Non-Saturating GAN}}\\
 \midrule
 Inf.& Tr.& Inf.& Tr.& 0.24& 0.89&31.48\\
 Inf.& Tr.& Tr.& Tr.& 0.25& 0.96&31.78\\
 Tr.& Tr.& Tr.& Tr.& 0.26& 1.04&29.46\\
 \bottomrule
    \end{tabular}
    \vspace{-0.1cm}
    \caption{(Top) Comparing different choices of noise distribution for the \textbf{Presto-S} process. (Bottom) for best performing noise distributions, performance for standard GAN design vs. proposed least-squares GAN.}
    \label{tab:edm_abl}
   \vspace{-0.3cm}
\end{wraptable}
\textbf{Loss Distribution Choice:} In~\tref{tab:edm_abl} (Top), we show the FAD, MMD, and CLAP score for many \textbf{Presto-S} distilled models with different noise distribution choices. 
We find that the original DMD2~\citep{yin2024improved} setup (first row) underperforms compared to adapting the loss distributions.
The largest change is in switching $p_{\text{\textcolor{dmd}{DMD}}}$ to the training distribution, which improves all metrics. 
This confirms our hypothesis that by focusing on the region most important for text guidance~\citep{kynkaanniemi2024applying}, we improve both audio quality and text adherence.
Switching $p_{\text{\textcolor{gan}{GAN}}}$ to the training distribution also helps; in this case, the discriminator is made to focus on higher-frequency features~\citep{Si2023FreeUFL}, benefiting quality. 
We also find only a small improvement when using the training distribution for $p_{\text{\textcolor{dsm}{DSM}}}$. 
This suggests that while the training distribution should lead to more stable learning of the online generator's score~\citep{wang2024evaluating}, this may not be crucial.
For all remaining experiments, we use $p_{\text{\textcolor{dmd}{DMD}}}(\sigma^{\text{train}}) = p_{\text{\textcolor{gan}{GAN}}}(\sigma^{\text{train}}) = p_{\text{\textcolor{dsm}{DSM}}}(\sigma^{\text{train}})$ 
and $p_{\text{\textcolor{gen}{gen}}}(\sigma^{\text{inf}})$.

\textbf{Discriminator Design:} We ablate the effect of switching from the chosen least-squares discriminator to the original softplus non-saturating discriminator, which notable treats the discriminator as a binary classifier and predicts the probability of real/generated. In~\tref{tab:edm_abl} (Bottom), we find that using the least-squares discriminator leads to consistent improvements in audio quality (FAD/MMD) and in particular text relevance (CLAP), owing to the increased stability from the least-squares GAN.
\begin{figure}[h]
    \vspace{-0.2cm}
    \centering
    \includegraphics[width=\linewidth]{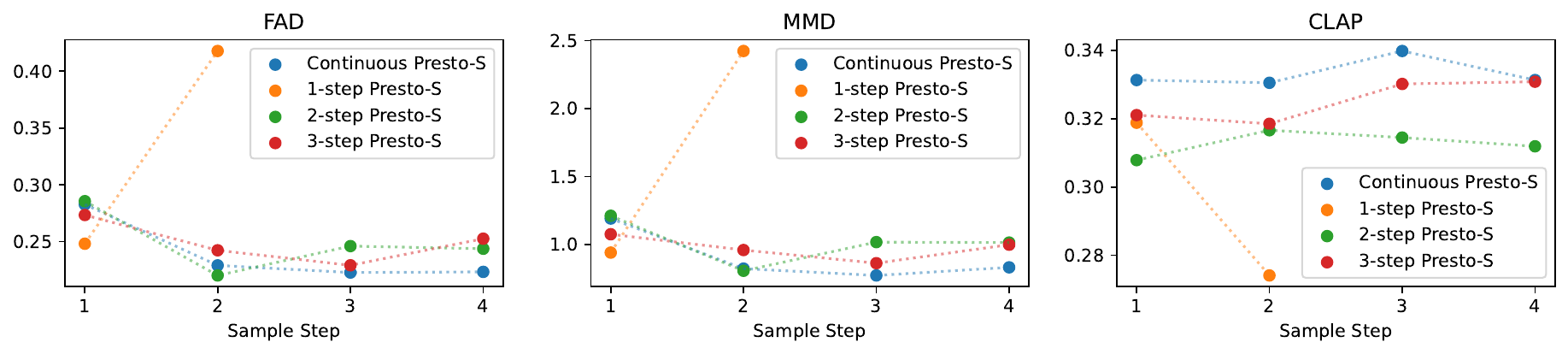}
    \vspace{-0.7cm}
    \caption{Continuous generator inputs vs. discrete inputs. Continuous inputs shows more consistent scaling with compute, while generally performing better in both quality and text relevance.}
    \label{fig:cont}
\end{figure}

\textbf{Continuous vs. Discrete Generator Inputs:}
We test how \emph{continuous-time} conditioning compares against a discrete and find the former is preferred as shown in~\fref{fig:cont}. 
Continuous noise levels maintain a correlation where more steps  improve quality, while discrete time models are more inconsistent.
Additionally, the continuous-time conditioning performs best in text relevance. 
While the 1 and 2-step discrete models show slightly better FAD metrics than continuous on 1 and 2-step sampling, these models have a failure mode as shown in~\fref{fig:hifreq}:
2-step discrete models drop  high-frequency information and render transients (i.e.~drum hits) poorly for genres like R\&B or hip-hop.

\subsection{\textbf{Presto-L} Results}

\begin{figure}[t]
\vspace{-0.4cm}
    \centering
    \includegraphics[width=\textwidth]{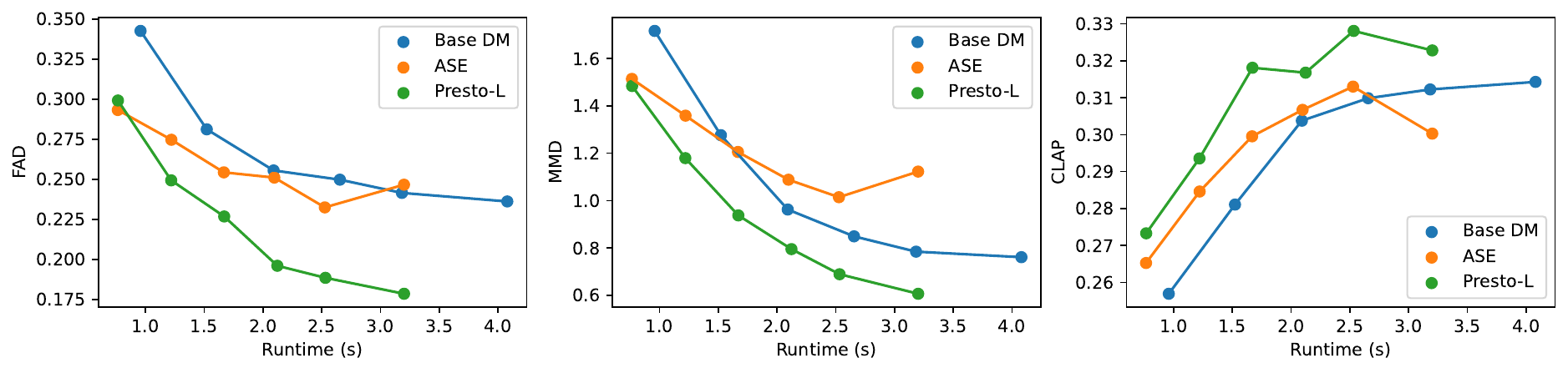}
    \vspace{-0.7cm}
    \caption{\textbf{Presto-L} results. \textbf{Presto-L} improves both the latency \emph{and} the overall performance across all metrics, against both the leading layer dropping baseline and the base model.}
    \label{fig:codda_results}
    \vspace{-0.4cm}
\end{figure}
We compare \textbf{Presto-L} with both our baseline diffusion model and ASE~\citep{Moon2024ASE} using the 2nd order DPM++ sampler~\citep{Lu2022DPMSolverFS} with CFG++~\citep{chung2024cfgpp}. 
For ASE and \textbf{Presto-L}, we use the optimal ``D3'' configuration from \citet{Moon2024ASE}, which corresponds to a dropping schedule, in terms of decreasing noise level (in quintiles), of $[14, 12, 8, 4, 0]$ (i.e.~we drop 14 layers for noise levels in the top quintile, 12 for the next highest quintile, and so on).
Layer distillation results at various sampling budgets are shown in~\fref{fig:codda_results}.
\textbf{Presto-L} yields an improvement over the base model on all metrics, speeding up by $\approx$27\% \emph{and} improving quality and text relevance. 
ASE provides similar acceleration but degrades performance at high sampling steps and scales inconsistently.
Dropping layers \emph{improving} performance can be viewed via the lens of multi-task learning, where (1) denoising each noise level is a different task (2) later layers only activating for lower noise levels enables specialization for higher frequencies. See \aref{app:presto_l_abl} for further ablations. %

\begin{table}[t]
    \tiny\centering
    \begin{tabular}{l c c | c c c c c c }
    \toprule
    Model  & NFE &RTF-M/S ($\uparrow$)& FAD ($\downarrow$) &MMD ($\downarrow$)&CLAP Score ($\uparrow$)  & Density ($\uparrow$) & Recall ($\uparrow$) & Coverage($\uparrow$)\\

     \midrule
        \multicolumn{5}{l}{\textbf{External Baselines*}} & &  & \\
    MusicGen-Small & 1.6K&0.77 &0.31&1.60&30.61&0.36&0.16&0.43\\
    MusicGen-Medium & 1.6K &0.39 &0.27&1.30&31.85&0.43&0.19&0.54\\
    MusicGen-Large &1.6K &0.37 &0.25&1.21&32.83&0.44&0.15&0.54\\
    Stable Audio Open&100&4.54 &0.23 &1.07&35.05&0.29&0.37&0.49\\

        \midrule
    \multicolumn{5}{l}{\textbf{Base Model, Diffusion Sampling}}& &  & \\
    DPM-2S & 80 & 7.72 / 7.34 & 0.24&  0.82&  31.56&  0.31 & 0.20 &  0.41\\
    DPM-2S+ASE & 80&  9.80 / 9.22&  0.25&  1.12&   30.03&  0.27&  0.16&   0.41\\
    DPM-2S+\textbf{Presto-L} (ours) & 80& 9.80 / 9.22& 0.18 & 0.61 &   32.28&  0.38& 0.29&   0.51\\
    
    \midrule
    \multicolumn{5}{l}{\textbf{Consistency-Based Distillation}} &  &  & \\
    CM& 4&  118.77  / 67.41&  0.47&    2.50 &  26.33 &   0.17&  0.01&  0.16\\
    SoundCTM &4 & 105.78 / 63.01& 0.35 &  1.72&  29.61& 0.17&  0.17 &  0.26\\
    DITTO-CTM & 4& 118.77  / 67.41&  0.36&    1.62& 28.31&  0.22&  0.04&   0.32\\
    \midrule
    \multicolumn{5}{l}{\textbf{Adversarial Distillation}} &  &  & \\
    DMD-GAN & 4& 118.77  / 67.41&   0.29 &   1.16 & 27.56 &  0.57& 0.07&   0.41\\
    \textbf{Presto-S} (ours) & 4&  118.77  / 67.41&  0.22&   0.83&  33.13&  0.60&  0.10 &  0.50\\
    \textbf{Presto-LS} (ours)& 4&  138.84 / 73.43 & 0.23  & 0.73 & 32.21&0.49  &  0.14 & 0.48  \\
    \bottomrule
    \end{tabular}
    \caption{Full Results on Song Describer (No vocals).$^*$External baseline RTFs are all natively stereo.}
    \label{tab:mainres}
    \vspace{-0.5cm}
\end{table}
\subsection{Full Comparison}
In~\tref{tab:mainres}, we compare against multiple baselines and external models.
For step distillation, \textbf{Presto-S} is best-in-class and the only distillation method to close to base model quality, while achieving an over 15x speedup in RTF from the base model.
Additionally, \textbf{Presto-LS} improves performance for MMD, beating the base model with further speedups (230/435ms latency for 32 second mono/stereo 44.1kHz on an A100 40 GB). We also find \textbf{Presto-LS}  improves  \emph{diversity} with higher recall. Overall, \textbf{Presto-LS} is 15x faster than SAO. We investigate latency more in \aref{app:rtf}.

\subsection{Listening Test}\label{sec:listening_test_main}
We also conducted a subjective listening test to compare \textbf{Presto-LS} with our base model, the best non-adversarial distillation technique SoundCTM~\citep{Saito2024SoundCTMUS} distilled from our base model, and Stable Audio Open~\citep{evans2024open}. Users ($n=16$) were given 20 sets of examples generated from each model 
(randomly cut to 10s for brevity) using random prompts from Song Describer and asked to rate the musical quality, taking into account both fidelity and semantic text match between 0-100. We run multiple paired t-tests with Bonferroni correction and find \textbf{Presto-LS} rates highest against all baselines ($p<0.05$). We show additional plots in~\fref{fig:mushra}.

\subsection{Presto-LS Qualitative Analysis}\label{sec:lsd_res}

\begin{wrapfigure}{r}{0.45\textwidth}
	\vskip -0.2in
	\centering
	\includegraphics[width=\linewidth, trim={0cm 0cm 0cm 0cm}, clip]{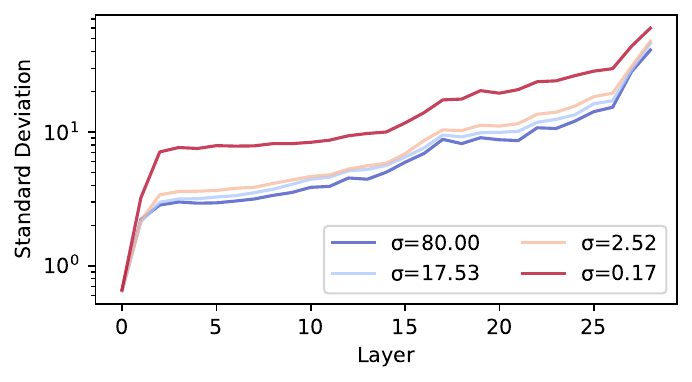}
 \vskip -0.15in
    \caption{\textbf{Presto-S} hidden activation var.}
	\label{fig:dmdsd}
\end{wrapfigure}

While \textbf{Presto-LS} improves speed and quality/diversity over step-only distillation, the increases are modest,
as the dropping schedule for \textbf{Presto-L} 
was reduced ($[12, 8, 8, 0, 0]$) for step distillation stability. 
To investigate more, we analyze the hidden state activation variance of our step-distilled model in~\fref{fig:dmdsd}.
The behavior is quite different than the base model,  
as the ``spike'' in the final layer is more amortized across the last 10 layers and never reaches the base model's magnitude. 
We hypothesize step-distilled models have more unique computation \emph{throughout} each DiT block, making layer dropping difficult. 

\subsection{Extensions}
\textbf{Adaptive Step Schedule:} A benefit of our continuous-time distillation is that besides setting how many steps (e.g., 1-4), we can set \emph{where} those steps occur along the diffusion process by tuning the $\rho$ parameter in the EDM inference schedule, which is normally set to $\rho = 7$. 
In particular, decreasing $\rho$ (lower bounded by 1) puts more weight on low-SNR features and increasing $\rho$ on higher-SNR features~\citep{karras2022elucidating}. 
Qualitatively, we find that this process enables increased diversity of outputs, even from the same latent code (see \aref{app:adapt_rho}).

\textbf{CPU Runtime:}
We benchmark \textbf{Presto-LS}'s speed performance for CPU inference.
On an Intel Xeon Platinum 8275CL CPU, we achieve a mono RTF of 0.74, generating 32 seconds of audio in 43.34 seconds. We hope to explore further CPU acceleration in future work.

\textbf{Fast Inference-Time Rejection Sampling:} Given \textbf{Presto-LS}'s speed,
we investigated using \emph{inference-time} compute to improve performance. 
Formally, we test the idea of \emph{rejection sampling}, inspired by \citet{Kim2023ConsistencyTM}, where we generate a batch of samples and reject $r$ fraction of them according to some ranking function. We use 
the CLAP score to discard samples that have poor text relevance. Over a number of rejection ratios (see \fref{fig:rej}), we find that CLAP rejection sampling strongly improves text relevance while maintaining or \emph{improving} quality at the cost of  diversity.

\section{Conclusion}
We proposed \textbf{Presto}, a dual-faceted approach to accelerating latent diffusion transformers by reducing sampling steps and cost per step via distillation. Our core contributions include the development of score-based distribution matching distillation (the first GAN-based distillation for TTM), a new layer distillation method, the first combined layer-step distillation, and evaluation showing each method are independently best-in-class and, when combined, can accelerate our base model by 10-18x (230/435ms latency for 32 second mono/stereo 44.1kHz, 15x faster than the comparable SOTA model), resulting in the fastest TTM model to our knowledge.  We hope our work will motivate continued work on (1) fusing step and layer distillation and (2) new distillation of methods for continuous-time score models across media modalities such as image and video.

\section*{Acknowledgements}
We would like to thank Juan-Pablo Caceres, Hanieh Deilamsalehy, and 
Chinmay Talegaonkar.

\section*{Ethics Statement and Reproducibility}
As TTM systems become more powerful, there is both the opportunity to increase accessibility of musical expression, but also concern such systems may compete with creators. 
To reduce risk, we train our TTM work only on instrumental \emph{licensed} music. 
Additionally, we hope that our focus on efficiency is useful to eventually make interactive-rate co-creation tools, allowing for greater flexibility and faster ideation. 
Following these concerns, we do not plan to release our model, but have done our best to compare against multiple open source baselines and/or re-train alternative methods for comparison and in-depth understanding of the reproducible insights of our work.

\bibliography{main}
\bibliographystyle{iclr2025_conference}
\clearpage
\appendix

\section{Appendix}

\subsection{Model Design Details}\label{app:arch}
As we perform latent diffusion, we first train a variational autoencoder. We build on the Improved RVQGAN~\citep{kumar2023high}  architecture and training scheme by using a KL-bottleneck with a dimension of 32 and an effective hop of 960 samples, resulting in an approximately 45 Hz VAE. We train to convergence using the recommended mel-reconstruction loss and the least-squares GAN formulation with L1 feature matching on multi-period and multi-band discriminators.

Our proposed base score model backbone builds upon DiT-XL~\citep{peebles2023scalable}, with modifications aimed at optimizing computational efficiency.
Specifically, we use a streamlined transformer block design, consisting of a single attention layer followed by a single feed-forward layer, similar to Llama~\citep{dubey2024llama}.
Our model utilizes three types of conditions including noise levels (timesteps) for score estimation, beat per minute (BPM) values of the song, and text descriptions.
Following EDM, we apply a logarithmic transformation to the noise levels, followed by sinusoidal embeddings.
Similarly, BPM values are input as scalars then go through sinusoidal embeddings to generate BPM embeddings. 
These processed noise-level embeddings and BPM embeddings are then combined and integrated into the DiT block through an adaptive layer normalization block.
For text conditioning, we compute text embedding tokens with T5-based encoders and concatenate with audio tokens at each attention layer. 
As a result, the audio token query attends to a concatenated sequence of audio and text keys, enabling the model to jointly extract relevant information from both modalities. To provide baseline architectural speedups, we use FlashAttention-2~\citep{dao2023flashattention} for the DiT and Pytorch 2.0's built in graph compilation~\citep{torch2} for the VAE decoder and MusicHifi mono-to-stereo.

\rebuttal{Our discriminator design follows \citet{yin2024improved} with a number of small modifications. $D_{\boldsymbol{\psi}}$ consists of 4 blocks of 1D convolutions interleaved with GroupNorm and SiLU activations, and a final linear layer to collapse the channel dimension. The discriminator thus does not use any final linear layer to project to a single value, and instead its' output is \emph{also} a 1D sequence but at even heavier downsampling than the input representation at $\approx$2.8 Hz. The discriminator receives its' input from the output of the 14th DiT Block (i.e.~the halfway point through our 28 block DiT), as DiTs lack a clear ``bottleneck'' layer to place the discriminator like in UNets. We leave further investigation into discriminator design and placement inside the model for future work.}

For the diffusion model hyparameter design, we follow \citet{karras2024analyzing}. Specifically, we set $\sigma_{\text{data}}=0.5$, $P_{\text{mean}}=-0.4$, $P_{\text{std}}=1.0$, $\sigma_{\text{max}}=80$, $\sigma_{\text{min}}=0.002$. We train the base model with $10\%$ condition dropout to enable CFG. The base model was trained for 5 days across 32 A100 GPUs with a batch size of 14 and learning rate of 1e-4 with Adam. For all score model experiments, we use CFG++~\citep{chung2024cfgpp} with $w=0.8$.

For \textbf{Presto-S}, following~\citet{yin2024improved} we use a fixed guidance scale of $w=4.5$ throughout distillation for the teacher model as CFG++ is not applicable for the distribution matching gradient. \rebuttal{We use 5 fake score model (and discriminator) updates per generator update,~following \citet{yin2024improved}, as we found little change in performance when varying the quantity around 5 (though using $\le3$ updates resulted in large training instability).} \rebuttal{Note that throughout \textbf{Presto-S}, the fake score model and the discriminator share an optimizer state.} Additionally, we use a learning rate of 5e-7 with Adam \rebuttal{for both the generator and fake score model / discriminator. We set $\nu_1 = 0.01$ and $\nu_2 = 0.005$ following \citet{yin2024improved}.} For all step distillation methods, we distill each model with a batch size of 80 across 16 Nvidia A100 GPUs for 32K iterations. We train all layer distillation methods for 60K iterations with a batch size of 12 across 16 A100 GPUs \rebuttal{with a learning rate of 8e-5. For \textbf{Presto-L}, we set $\nu=0.1$.} 

\subsection{Experimental Details}\label{app:bases}
\subsubsection{Baseline Details}
\rebuttal{Our benchmarks are divided into two main classes: acceleration algorithms and external open-source models. For acceleration algorithms, we distill our internal base model per method, utilizing publicly available code as a reference when available \citep{song2023consistency, Saito2024SoundCTMUS, yin2024improved}. For the open-source external models, we use the models directly in their default setups as recommended by \citet{copet2023simple, evans2024open}.}
\begin{itemize}
    \item Consistency Models (CM)~\citep{song2023consistency, bai2023accelerating}:
    This distillation technique learns a mapping from anywhere on the diffusion process to the data distribution (i.e.~$\bm{x}_t \rightarrow \bm{x}_0$) by enforcing the self-consistency property that $G_{\boldsymbol{\phi}}(\bm{x}_t, t) = G_{\boldsymbol{\phi}}(\bm{x}_{t'}, t') \quad \forall t, t'$. We follow the parameterization used in past audio works \citep{bai2023accelerating, Novack2024DITTO2DD} that additionally distills the CFG parameter into the model directly.
    \item SoundCTM~\citep{Saito2024SoundCTMUS}:
    This approach distills a model into a consistency \emph{trajectory} model \citep{Kim2023ConsistencyTM} that enforces the self-consistency property, learning an anywhere-to-anywhere mapping. SoundCTM forgoes the original CTM adversarial loss and calculates the consistency loss via intermediate base model features.
    
    \item DITTO-CTM~\citep{Novack2024DITTO2DD},
    This audio approach is also based off of \citep{Kim2023ConsistencyTM}, yet brings the consistency loss back into the raw outputs and instead replaces CTM's multi-step teacher distillation with single-step teacher (like CMs) and removes the learned target timestep embedding, thus  more efficient (though less complete) than SoundCTM.

    \item DMD-GAN~\citep{yin2024improved}:
    This approach removes the distribution matching loss from DMD2, making it a fully GAN-based finetuning method, which is in line with past adversarial distillation methods \citep{Sauer2023AdversarialDD}).

    \item ASE~\citep{Moon2024ASE}, 
    This funetuning approach for diffusion models, as discussed in Sec.~\ref{sec:codda}, finetunes the base model with the standard DSM loss, but for each noise level drops a fixed number of layers, starting at the back of the diffusion model's DiT blocks.

    \item MusicGen~\citep{copet2023simple}: MusicGen is a non-diffusion based music generation model that uses an autoregressive model to predict discrete audio tokens \citep{defossez2022highfi} at each timestep in sequence, and comes in small, medium, and large variants (all stereo).

    \item Stable Audio Open~\citep{evans2024open}: Stable Audio Open is a SOTA open-source audio diffusion model, which can generate variable lengths up to 45s in duration. Stable Audio Open follows a similar design to our base model, yet uses cross-attention for conditioning rather than AdaLN which we use, which increases runtime.
\end{itemize}

\subsubsection{Metrics Details}
We use Frechet Audio Distance (FAD)~\citep{kilgour2018fr}, Maximum Mean Discrepancy (MMD) \citep{jayasumana2024rethinking}, and Contrastive Language-Audio Pretraining (CLAP) score~\citep{wu2023large}, all with the CLAP-LAION music backbone~\citep{wu2023large} given its high correlation with human perception~\citep{fadtk}. FAD and MMD measure audio quality/realness with respect to Song Describer (lower better), and CLAP score measures prompt adherence (higher better). 
When comparing to other models, we 
also include density (measuring quality), recall and coverage (measuring diversity)~\citep{naeem2020reliable}, and real-time factor (RTF) for both mono (M) and stereo (S, using MusicHiFi), which measures the total seconds of audio generated divided by the generation time, where higher is better for all.

\clearpage
\subsection{\textbf{Presto-S} Algorithm}\label{app:presto_s_algo}

\begin{algorithm}[h!]
    \caption{\textbf{Presto-S}}
    \label{alg:edm3}
    \begin{algorithmic}[1]
    \INPUT: generator $G_{\boldsymbol{\phi}}$, real score model $\mu_{\text{real}}$, fake score model $\mu_{\boldsymbol{\psi}}$, discriminator $D_{\boldsymbol{\psi}}$ , CFG weight $w$, $p_{\text{\textcolor{gen}{gen}}}(\sigma^{\text{inf}})$, $p_{\text{\textcolor{dmd}{DMD}}}(\sigma^{\text{train}})$, $p_{\text{\textcolor{dsm}{DSM}}}(\sigma^{\text{train}})$, $p_{\text{\textcolor{gan}{GAN}}}(\sigma^{\text{train}})$, real sample $\bm{x}_{\text{real}}$, GAN weights $\nu_1, \nu_2$, optimizers $g_1, g_2$, weighting function $\lambda$
        \STATE $\textcolor{gen}{\sigma} \sim p_{\text{\textcolor{gen}{gen}}}(\sigma^{\text{inf}})$
        \STATE $\epsilon_{\text{gen}} \sim \mathcal{N}(0, \bm{I})$
        \STATE $\textcolor{gout}{\hat{\bm{x}}_{\text{gen}}} = G_{\boldsymbol{\phi}}(\bm{x}_{\text{real}} + \textcolor{gen}{\sigma}\epsilon_{\text{gen}}, \textcolor{gen}{\sigma})$
        \IF{generator turn}
            \STATE $\textcolor{dmd}{\sigma} \sim p_{\text{\textcolor{dmd}{DMD}}}(\sigma^{\text{train}})$
            \STATE $\epsilon_{\text{dmd}} \sim \mathcal{N}(0, \bm{I})$
            \STATE $\nabla_{\boldsymbol{\phi}} \mathcal{L}_{\text{DMD}} =\left((\mu_{\boldsymbol{\psi}}(\textcolor{gout}{\hat{\bm{x}}_{\text{gen}}} + \textcolor{dmd}{\sigma}\epsilon_{\text{dmd}}, \textcolor{dmd}{\sigma}) - \tilde\mu^w_{\text{real}}(\textcolor{gout}{\hat{\bm{x}}_{\text{gen}}} + \textcolor{dmd}{\sigma}\epsilon_{\text{dmd}}, \textcolor{dmd}{\sigma})\right) \cdot \nabla_{\boldsymbol{\phi}}\textcolor{gout}{\textcolor{gout}{\hat{\bm{x}}_{\text{gen}}}}$
            \STATE $\textcolor{gan}{\sigma} \sim p_{\text{\textcolor{gan}{GAN}}}(\sigma^{\text{train}})$
            \STATE $\epsilon_{\text{fake}} \sim \mathcal{N}(0, \bm{I})$
            \STATE $\mathcal{L}_{\text{GAN}} = \|1 - D_{\boldsymbol{\psi}}(\textcolor{gout}{\hat{\bm{x}}_{\text{gen}}} + \textcolor{gan}{\sigma}\epsilon_{\text{fake}}, \textcolor{gan}{\sigma})\|_2^2$
            \STATE $\boldsymbol{\phi} \leftarrow \boldsymbol{\phi}- g_1(\nabla_{\boldsymbol{\phi}} \mathcal{L}_{\text{DMD}} + \nu_1 \nabla_{\boldsymbol{\phi}} \mathcal{L}_{\text{GAN}})$
        \ELSE
            \STATE $\textcolor{dsm}{\sigma} \sim p_{\text{\textcolor{dsm}{DSM}}}(\sigma^{\text{train}})$
            \STATE $\epsilon_{\text{dsm}} \sim \mathcal{N}(0, \bm{I})$
            \STATE $\mathcal{L}_{\text{fake-DSM}} = \lambda(\textcolor{dsm}{\sigma})\|\textcolor{gout}{\hat{\bm{x}}_{\text{gen}}} - \mu_{\boldsymbol{\psi}}(\textcolor{gout}{\hat{\bm{x}}_{\text{gen}}} + \textcolor{dsm}{\sigma}\epsilon_{\text{dsm}}, \textcolor{dsm}{\sigma})\|_2^2$
            \STATE $\textcolor{gan}{\sigma}_{\text{real}}, \textcolor{gan}{\sigma}_{\text{fake}} \sim p_{\text{\textcolor{gan}{GAN}}}(\sigma^{\text{train}})$
            \\\STATE $\epsilon_{\text{real}},  \epsilon_{\text{fake}} \sim \mathcal{N}(0, \bm{I})$
            \STATE $\mathcal{L}_{\text{GAN}} = \|D_{\boldsymbol{\psi}}(\textcolor{gout}{\hat{\bm{x}}_{\text{gen}}} + \textcolor{gan}{\sigma}_{\text{fake}}\epsilon_{\text{fake}}, \textcolor{gan}{\sigma}_{\text{fake}})\|_2^2 + \|1 - D_{\boldsymbol{\psi}}(\bm{x}_{\text{real}} + \textcolor{gan}{\sigma}_{\text{real}}\epsilon_{\text{real}}, \textcolor{gan}{\sigma}_{\text{real}})\|_2^2$
            \STATE $\boldsymbol{\psi} \leftarrow \boldsymbol{\psi}- g_2(\nabla_{\boldsymbol{\psi}} \mathcal{L}_{\text{fake-DSM}} + \nu_2 \nabla_{\boldsymbol{\psi}} \mathcal{L}_{\text{GAN}})$
        \ENDIF
    \OUTPUT: $\boldsymbol{\phi}, \boldsymbol{\psi}$
\end{algorithmic}
\end{algorithm}
We outline a condensed algorithm of \textbf{Presto-S} in math notation in  Algorithm~\ref{alg:edm3}.

\subsection{\rebuttal{\textbf{Presto-S} Pseudo-code Walkthrough}\label{app:walkthrough}}
\rebuttal{We provide a comprehensive algorithm walkthrough using PyTorch psuedo-code of our \textbf{Presto-S} training loop below. To perform \textbf{Presto-S}, we first define the corruption process for any given clean sample, according to either the training $p(\sigma^{\text{train}})$ or the inference $p(\sigma^{\text{inf}})$ noise distribution:}
\begin{lstlisting}[language=Python]
def diffuse(x, dist):
  eps = noise_normal_like(x)
  if dist == 'training':
    sigma = training_dist_like(x)
  elif dist == 'inference':
    sigma = inference_dist_like(x)
  return x + sigma * eps, sigma
\end{lstlisting}
\rebuttal{We then define each of the component loss functions for the \textbf{Presto-S} continuous-time DMD2 distillation process. This corresponds to the three loss types: the distribution matching loss, the least-squares GAN loss, and the fake denoising score matching loss. For the distribution matching loss, we corrupt some generated sample according to the training distribution and then pass that into both the fake and real score models (where the real score model uses classifier-free guidance). The difference in these scores forms the distribution matching gradient:}
\begin{lstlisting}[language=Python]
def dmd(x, real_score_model, fake_score_model, cfg):
  x_noise, sigma = diffuse(x, 'training')
  fake_denoised = fake_score_model(x_noise, sigma)
  real_denoised = real_score_model(x_noise, sigma, cfg)
  return fake_denoised - real_denoised
\end{lstlisting}
\rebuttal{For the least-squares GAN loss, we corrupt some sample (either real or generated) according to the training distribution and pass this through the discriminator (which itself involves first passing through some of the fake score model to extract intermediate features). The output of the discriminator is then passed into the least-squares loss against some target value (i.e.~the generator wants to push the discriminator outputs on generated samples towards 1, while the discriminator aims to push generated samples towards 0 and real samples towards 1):}
\begin{lstlisting}[language=Python]
def gan(x, discriminator, tgt=1):
  x_noise, sigma = diffuse(x, 'training')
  d_out = discriminator(x_noise, sigma)
  return mse(tgt, d_out)
\end{lstlisting}
\rebuttal{Finally, we have the fake DSM loss. This loss is identical to the normal diffusion loss (with a weighted MSE between the outputs of the score model and the clean data), yet will be calculated treating \emph{generator} outputs as the ground truth clean data and using the fake score model:}
\begin{lstlisting}[language=Python]
def dsm(x, fake_score_model):
  x_noise, sigma = diffuse(x, 'training')
  x_denoised = fake_score_model(x_noise, sigma)
  return weighted_mse(x, x_denoised, sigma)
\end{lstlisting}

\rebuttal{Given these helper loss functions, we can now proceed with the main distillation loop, which is as follows. For both the generator and discriminator turns, we first corrupt some real input data according to the inference distribution, and pass this through our generator to get the generator outputs \texttt{x\_denoised} (steps (1) and (4) in \fref{fig:presto_s_ext_train}). If it is a generator turn (which happens once for every 5 fake score turns), we calculate the distribution matching loss (step (2)) and the generator adversarial loss (step (3)) on \texttt{x\_denoised} and update the generator. If it is a fake score turn, we calculate and the discriminator's adversarial loss (step (5)) on both the generated \texttt{x\_denoised} and real samples \texttt{x} and the fake DSM loss (step (6)) on \texttt{x\_denoised}, thus updating the fake score model and the discriminator:}

\begin{lstlisting}[language=Python]
def forward(
  x, generator, discriminator, fake_score_model, real_score_model, generator_turn, nu_1, nu_2
):
  # step (1) and (4)
  x_noise, sigma = diffuse(x, 'inference')
  x_denoised = generator(x_noise, sigma)

  if generator_turn: # GENERATOR TURN
    # Distribution Matching Loss, step (2)
    dmd_loss = dmd(x_denoised, real_score_model, fake_score_model, cfg)

    # Generator Adversarial Loss, step (3)
    g_loss = gan(x_denoised, discriminator, 1)

    loss = dmd_loss + nu_1 * g_loss
  else: # FAKE SCORE TURN
    # Discriminator Adversarial Loss, step (5)
    d_loss = gan(x, discriminator, 1) + gan(x_denoised, discriminator, 0)
    
    # fake DSM loss, step (6)
    dsm_loss = dsm(x_denoised, fake_score_model)

    loss = dsm_loss + nu_2 * d_loss
  return loss
\end{lstlisting}
\rebuttal{This constitutes one full update of the \textbf{Presto-S} process, alternating between the generator and fake score model / discriminator updates. At inference time, we can feed in pure noise and alternate between generating clean data with our generator and adding progressively smaller noise back to the generation (for some pre-defined list of noise levels), allowing for multi-step sampling:}
\clearpage
\begin{lstlisting}[language=Python]
def inference(generator, sigmas, start_noise):
  x = start_noise
  for sigma in sigmas:
    x = x + noise_normal_like(x) * sigma
    x = generator(x, sigma)
  return x
\end{lstlisting}

\subsection{\rebuttal{\textbf{Presto-S} Expanded Diagram}\label{app:expanded_diagram}}
\begin{figure}[h!]
    \centering
    \includegraphics[width=\linewidth,trim={0cm 5.20cm 0 0},clip]{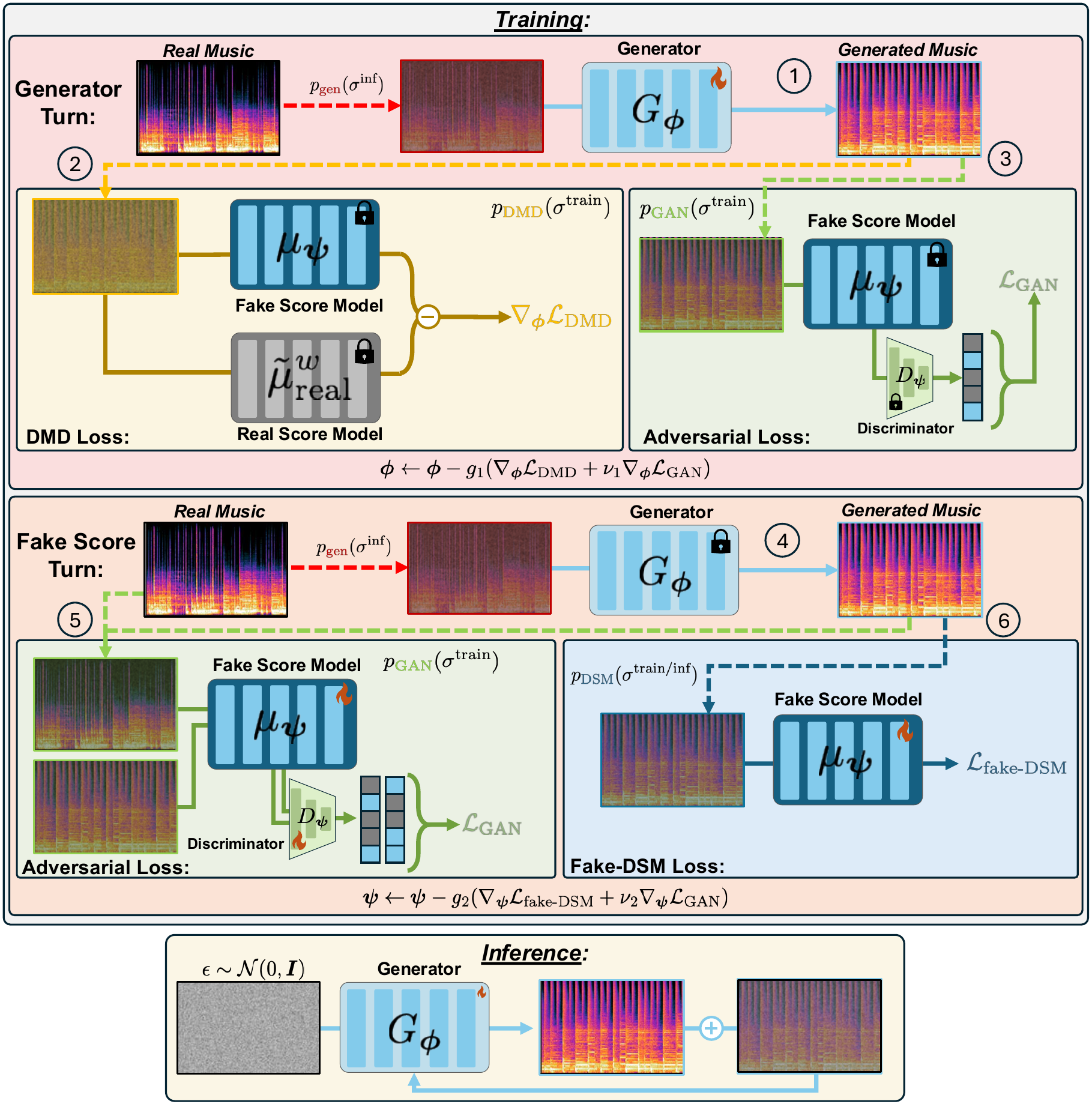}
    \rebuttal{\caption{Presto-S training process.
    }\label{fig:presto_s_ext_train}}
\end{figure}

For an in-depth visual illustration of \textbf{Presto-S}, please see ~\fref{fig:presto_s_ext_train} and~\fref{fig:presto_s_ext_inf} for expanded training and inference diagrams. 

\begin{figure}[h!]
    \centering
    \includegraphics[width=\linewidth,trim={0cm 0.0cm 0 26.2cm},clip]{imgs/presto_fig_extended.pdf}
    \rebuttal{\caption{Presto-S inference. For multi-step sampling, we use ping-pong-like sampling.
    }\label{fig:presto_s_ext_inf}}
\end{figure}

\newpage
\subsection{\rebuttal{\textbf{Presto-L} Algorithm}}\label{app:presto_l_alg}
\begin{algorithm}[h!]
    \caption{\rebuttal{\textbf{Presto-L}}}
    \label{alg:presto_l}
    \begin{algorithmic}[1]
    \INPUT: \rebuttal{pre-trained score model $\mu_{\boldsymbol{\theta}}$, real sample $\bm{x}_{\text{real}}$, self-teacher weight $\nu$, optimizer $g, g_2$, weighting function $\lambda$, \# of DiT blocks $B$, budget mapping $\ell$, layer drop function $\mathbf{LD}$}
        \STATE $\sigma \sim p(\sigma^{\text{train}})$
        \STATE $b = \ell(\sigma)$
        \STATE $\epsilon \sim \mathcal{N}(0, \bm{I})$
        \STATE $\hat{\bm{x}}_{\bm{L}}, h_{\bm{L}} = \mathbf{LD}(\mu_{\boldsymbol{\theta}}, \bm{x}_{\text{real}} + \sigma\epsilon, \sigma, b)$
        \STATE $\hat{\bm{x}}_{\text{full}}, h_{\text{full}} = \mathbf{LD}(\mu_{\boldsymbol{\theta}}, \bm{x}_{\text{real}} + \sigma\epsilon, \sigma, B)$
        \STATE $\mathcal{L}_{\text{DSM}} = \lambda(\sigma)\|\bm{x}_{\text{real}} - \hat{\bm{x}}_{\bm{L}}\|_2^2$
        \STATE $\mathcal{L}_{\text{st}} = \|h_{\bm{L}} - \texttt{sg}(h_{\text{full}})\|_2^2$
        \STATE $\boldsymbol{\theta} \leftarrow \boldsymbol{\theta}- g(\nabla_{\boldsymbol{\theta}} \mathcal{L}_{\text{DSM}} + \nu \nabla_{\boldsymbol{\theta}} \mathcal{L}_{\text{st}})$
    \OUTPUT: $\boldsymbol{\theta}$
\end{algorithmic}
\end{algorithm}
\begin{algorithm}[h!]
    \caption{\rebuttal{$\mathbf{LD}$: Modified DiT forward pass with layer dropping and budget conditioning.}}
    \label{alg:LD}
    \begin{algorithmic}[1]
    \INPUT: \rebuttal{score model noise embedder $\mu_{\boldsymbol{\theta}}^\text{noise}$, score model budget embedder $\mu_{\boldsymbol{\theta}}^\text{budget}$, score model DiT blocks $\{\mu_{\boldsymbol{\theta}}^i\}_{i=1}^B$, score model budget AdaLN $\mu_{\boldsymbol{\theta}}^{\text{LN}}$, score model output layer $\mu_{\boldsymbol{\theta}}^{\text{final}}$, input $\bm{x}$, noise level $\sigma$, budget $b$}
        \STATE $\bm{e}_\sigma = \mu_{\boldsymbol{\theta}}^\text{noise}(\sigma)$ \hfill \texttt{\footnotesize{// embed noise level}}
        \STATE $\bm{e}_b = \mu_{\boldsymbol{\theta}}^\text{budget}(b)$ \hfill \texttt{\footnotesize{// embed budget}}
        \STATE $\bm{e} = \bm{e}_\sigma + \bm{e}_b$
        \FOR{ $i:= 1$ to $b-1$} 
        \STATE \texttt{\footnotesize{// apply first b-1 DiT blocks}}
            \STATE $x = \mu_{\boldsymbol{\theta}}^i(x, \bm{e})$
        \ENDFOR
        \STATE $x = \mu_{\boldsymbol{\theta}}^B(x, \bm{e})$ \hfill \texttt{\footnotesize{// apply final DiT block}}
        \STATE $x = \mu_{\boldsymbol{\theta}}^{\text{LN}}(x, \bm{e}_b)$ \hfill \texttt{\footnotesize{// apply budget-based AdaLN}}
        \STATE $h = x/\|x\|_2$ \hfill \texttt{\footnotesize{// get normalized hidden state for $\mathcal{L}_{\text{st}}$}}
    \OUTPUT: $\mu_{\boldsymbol{\theta}}^{\text{final}}(x), h$
\end{algorithmic}
\end{algorithm}

\rebuttal{We show the full algorithm in detail for Presto-L in Algorithm~\ref{alg:presto_l}, which proceeds as a modified version of standard diffusion training like in \citet{Moon2024ASE}. We first sample some noise level $\sigma$, and then map the noise level to its corresponding budget $b$ given some mapping function $\ell(\cdot)$. Following \citet{Moon2024ASE}, $\ell: \mathbb{R} \rightarrow \{i\}_{i=1}^B$ is a deterministic map from the percentile of the noise level according to the training noise distribution $F(\sigma)$ (where $F$ is the cumulative distribution function) to some budget amount, which we write as $[q_1, q_2, q_3, q_4, q_5]$ for a mapping based on descending \emph{quintiles} (e.g. $q_1=14$ means that all noise levels in the largest quintile drop 14 layers).}

\rebuttal{We then call the modified forward function of the model $\mathbf{LD}$ (see Algorithm~\ref{alg:LD}) on the noisy inputs with both the given budget $b$ and the full budget $B$ (i.e. using all DiT blocks). $\mathbf{LD}$ modifies the forward pass of the model by (1) adding a global budget embedding that is added to the noise embedding (2) only iterating through the first $b-1$ DiT blocks followed by the final DiT block (to preserve final block behavior, see \sref{sec:codda}) (3) adding an additional AdaLN conditional only on the budget after the final DiT block and (4) also returning the normalized hidden state of the model (i.e.~the input to the final layer of the DiT, normalized along the channel dimension). We calculate the standard denoising score matching loss $\mathcal{L}_{\text{DSM}}$ as normal but with our layer-dropped outputs, and additionally calculate $\mathcal{L}_{\text{st}}$ as the MSE between the layer-dropped hidden state and the full budget hidden state (with a stop-gradient operation on the full budget pass.}

\subsection{Analyzing Failure Modes of Combined Layer and Step Distillation}\label{app:lsfails}
We empirically discovered a number of failure modes when trying to combine step and layer distillation. As noted in~\sref{sec:lsd_res}, the heavier per-layer requirements of distilled few-step generation made all standard dropping schedules~\citep{Moon2024ASE} intractable and prone to quick generator collapse, necessitating a more conservative dropping schedule. In~\fref{fig:lsfails}, we show the generator loss, discriminator loss, distribution matching gradient, and the discriminator's accuracy for the \emph{real} inputs over distillation, for a number of different setups:
\begin{figure}[h]
    \centering
    \includegraphics[width=.8\textwidth]{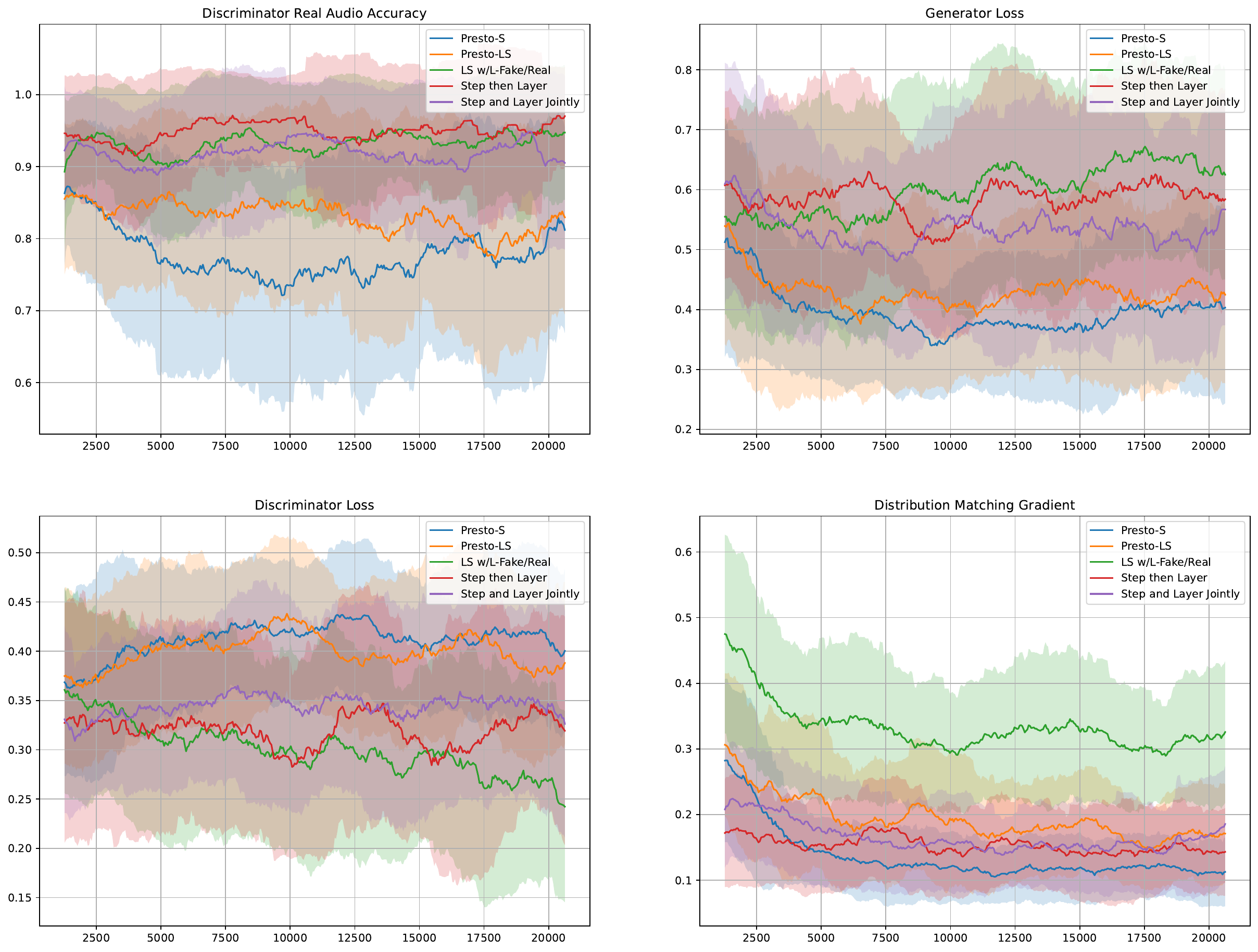}
    \caption{Step distillation losses for early distillation for multiple combination methods. \textbf{Presto-LS} is the only setup that avoids generator degradation and high variance distribution matching gradients.}
    \vspace{-.35cm}
    \label{fig:lsfails}
\end{figure}
\begin{itemize}
    \item \textbf{Presto-S}, pure step distillation mechanism (blue).
    \item \textbf{Presto-LS}, optimal combined setup where we pretrain the model with \textbf{Presto-L} and then perform \textbf{Presto-S}, but with keeping the real and fake score models initialized from the original score model (orange).
    \item LS with L-Fake/Real, which mimics \textbf{Presto-LS} but uses the \textbf{Presto-L} model for the fake and real score models as well (green).
    \item Step then Layer, where we first perform \textbf{Presto-S} distillation and then continue distillation with \textbf{Presto-L} layer dropping on the generator (red).
    \item Step and Layer jointly, where we perform \textbf{Presto-S} and \textbf{Presto-L} at the same time initialized from the original score model (purple),
\end{itemize}
We see that the runs which do not initialize with pretrained \textbf{Presto-L} (Step then Layer, Step and Layer) show clear signs of generator degradation, with increased generator loss, decreased discriminator loss, and notably near perfect accuracy on real samples, as attempting to learn to drop layers from scratch during step distillation gives strong signal to the discriminator. Additionally,  LS with L-Fake/Real inherits similar collapse issues but has a higher variance distribution matching gradient as the layer-distilled real and fake score models are poor estimators of the gradient.

\subsection{Inference-time noise schedule Sensitivity analysis}\label{app:adapt_rho}
Given our final \textbf{Presto-LS} distilled 4-step generator, we show how changing the inference-time noise schedule can noticeably alter the outputs, motivating our idea of a continuous-time conditioning.

\begin{figure}[h!]
    \centering
    \vspace{-0.5cm}
    \includegraphics[width=\textwidth]{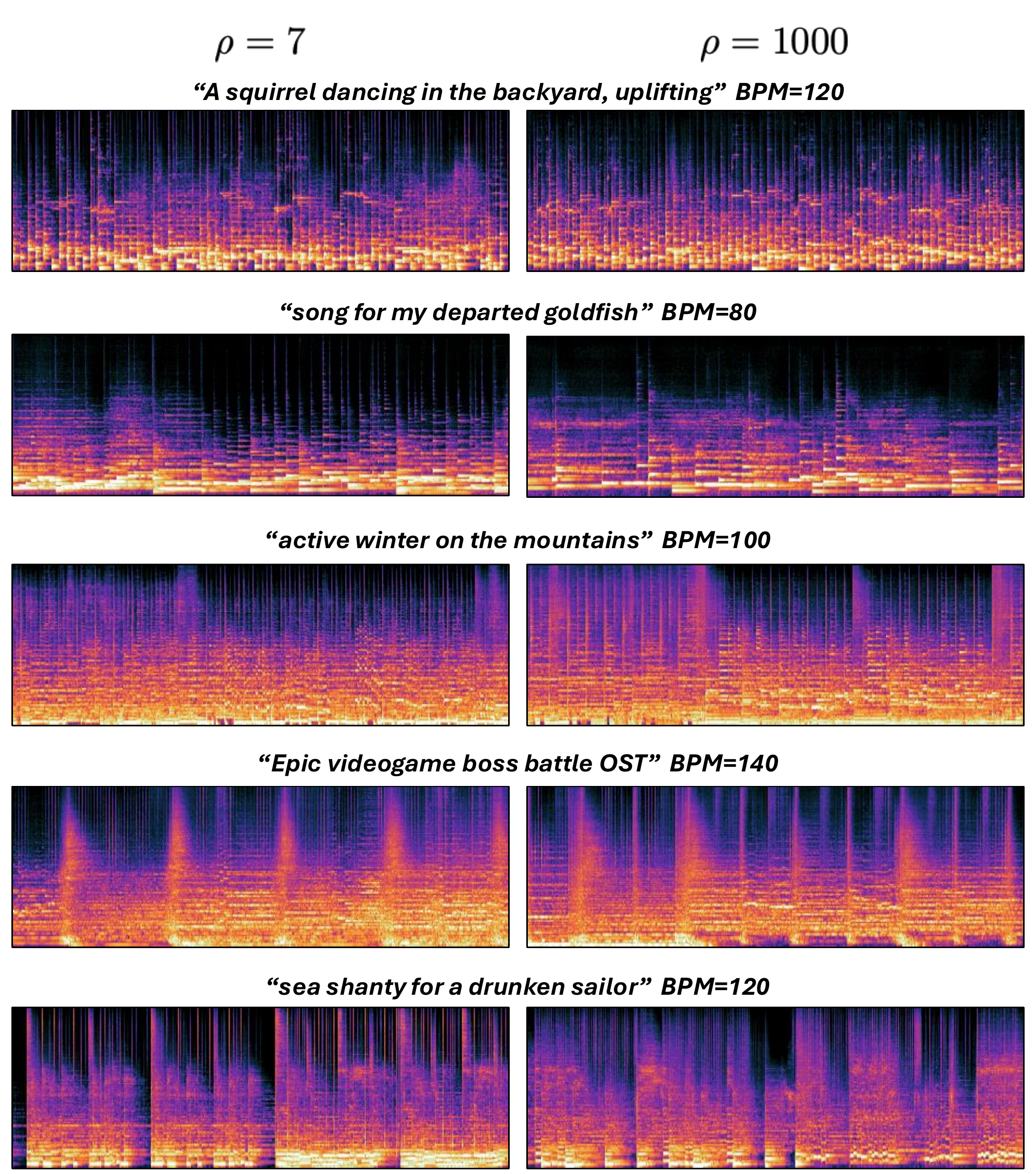}
    \vspace{-0.5cm}
    \caption{Generations from \textbf{Presto-LS} from the \emph{same} text prompt and latent code (i.e. starting noise and added noise during sampling), only varying the $\rho$ parameter between (7 and 1000). Purely shifting the noise schedule for 4-step sampling allows for perceptually distinct outputs.}
    \label{fig:adapt_rho}
    \vspace{-.5cm}
\end{figure}

The EDM inference schedule follows the form of:
\begin{equation}
    \sigma_{i < N} = \left(\sigma_{\text{max}}^{1/\rho} + \frac{i}{N-1} (\sigma_{\text{min}}^{1/\rho} - \sigma_{\text{max}}^{1/\rho})\right)^{\rho},
\end{equation}
where increasing the $\rho$ parameter puts more weight on the low-noise, high-SNR regions of the diffusion process. In \fref{fig:adapt_rho}, we show a number of samples generated from \textbf{Presto-LS} with identical conditions and latent codes (i.e. starting noise and all other added gaussian noise during sampling), only changing $\rho$, from the standard of 7 to 1000 (high weight in low-noise region). We expect further inference-time tuning of the noise schedule to be beneficial.

\subsection{RTF Analysis}\label{app:rtf}
We define the RTF for a model $\boldsymbol{\theta}$ as:
$\text{RTF}_b(\boldsymbol{\theta}) = \frac{ b T_{\boldsymbol{\theta}}}{\text{latency}_{\boldsymbol{\theta}}(b)}$, where $T_{\boldsymbol{\theta}}$ is the generation duration or how much \emph{contiguous} audio the model can generate at once
and $\text{latency}_{\boldsymbol{\theta}}(b)$ is the time it takes for generation following \citep{Evans2024LongformMG, Zhu2024MusicHiFiFH}.
This is different from the fixed-duration batched RTF used in \citet{nistal2024diff}. 
We test $b=1$ as well as the \emph{maximum} batch size we could attain for each model on a single A100 40GB to get a sense of maximum throughput. 
We show results in \tref{tab:rtf1} and \tref{tab:rtfmax} for all components of our generative process, including latency metrics for generation (i.e.~the diffusion model or distilled generator), decoding (i.e.~VAE decoder from latents to audio) and the optional mono-to-stereo (M2S), as well as overall RTF/latency for mono and stereo inference. 
We omit the MusicGen models and the other step-distillation methods which share the same RTF as \textbf{Presto-S}. 
For the fastest model \textbf{Presto-LS}, the biggest latency bottleneck is the mono-to-stereo model~\citep{Zhu2024MusicHiFiFH} and VAE decoder.
In future work, we hope to optimize the VAE and mono-to-stereo modules for faster inference.

\begin{table}
    \centering
    \small
    \begin{tabular}{l|cccc|ccc}
    \toprule
         \multirow{2}{*}{Model}&  Generation &  Decoding &  Mono&   Mono&  M2S &  Stereo &  Stereo \\
 & Latency  & Latency & RTF& Latency & Latency & RTF&Latency \\
 \midrule
  Stable Audio Open& 6159.01& 887.99& N/A& N/A& 0& 4.54&7047\\
         Base DM&  4079.81&  64.45&  7.72&  4144.27&  205.31&  7.36& 4349.58\\
         ASE&  3200.73&  64.45&  9.80&  3265.19&  205.31&  9.22& 3470.50\\
         Presto-L&  3201.19&  64.45&  9.80&  3265.64&  205.31&  9.22& 3470.95\\
 SoundCTM& 238.06& 64.45& 105.78& 302.51& 205.31& 63.01&507.83\\
 Presto-S& 204.98& 64.45& 118.77& 269.43& 205.31& 67.41&474.74\\
 Presto-LS& 166.04& 64.45& 138.84& 230.49& 205.31& 73.43&435.8\\
 \bottomrule
    \end{tabular}
    \caption{Latency (ms) and real-time factor for a batch size of one on an A100 40GB GPU.}
    \label{tab:rtf1}
\end{table}

\begin{table}
    \centering
    \small
    \begin{tabular}{l|cccc|ccc}
    \toprule
         \multirow{2}{*}{Model}&  Generation &  Decoding &  Mono&   Mono&  M2S &  Stereo &  Stereo \\
 & Latency & Latency & RTF& Latency & Latency & RTF&Latency \\
 \midrule
         Stable Audio Open&   34602.86&  4227.54&  N/A&  N/A&  0&  7.42& 38830.4\\
 Base& 18935.26& 1198.21& 14.3& 20133.46& 1775.73& 96.38&21909.19\\
 ASE& 14584.85& 1198.21& 18.25& 15783.05& 1775.73& 96.25&17558.78\\
 Presto-L& 14655.02& 1198.21& 18.17& 15853.23& 1775.73& 96.25&17628.96\\
 SoundCTM& 1135.65& 1198.21& 123.4& 2333.86& 1775.73& 92.98&4109.58\\
 Presto-S& 715.41& 1198.21& 150.5& 1913.62& 1775.73& 92.18&3689.34\\
 Presto-LS& 695.19& 1198.21& 152.11& 1893.4& 1775.73& 92.13&3669.13\\
 \bottomrule
    \end{tabular}
    \caption{Latency (ms) and real-time factor for max batch size on an A100 40GB GPU.}
    \label{tab:rtfmax}
\end{table}

\subsection{Presto-L Design Ablation}\label{app:presto_l_abl}

\begin{figure}[h!]
    \centering
    \includegraphics[width=.9\textwidth]{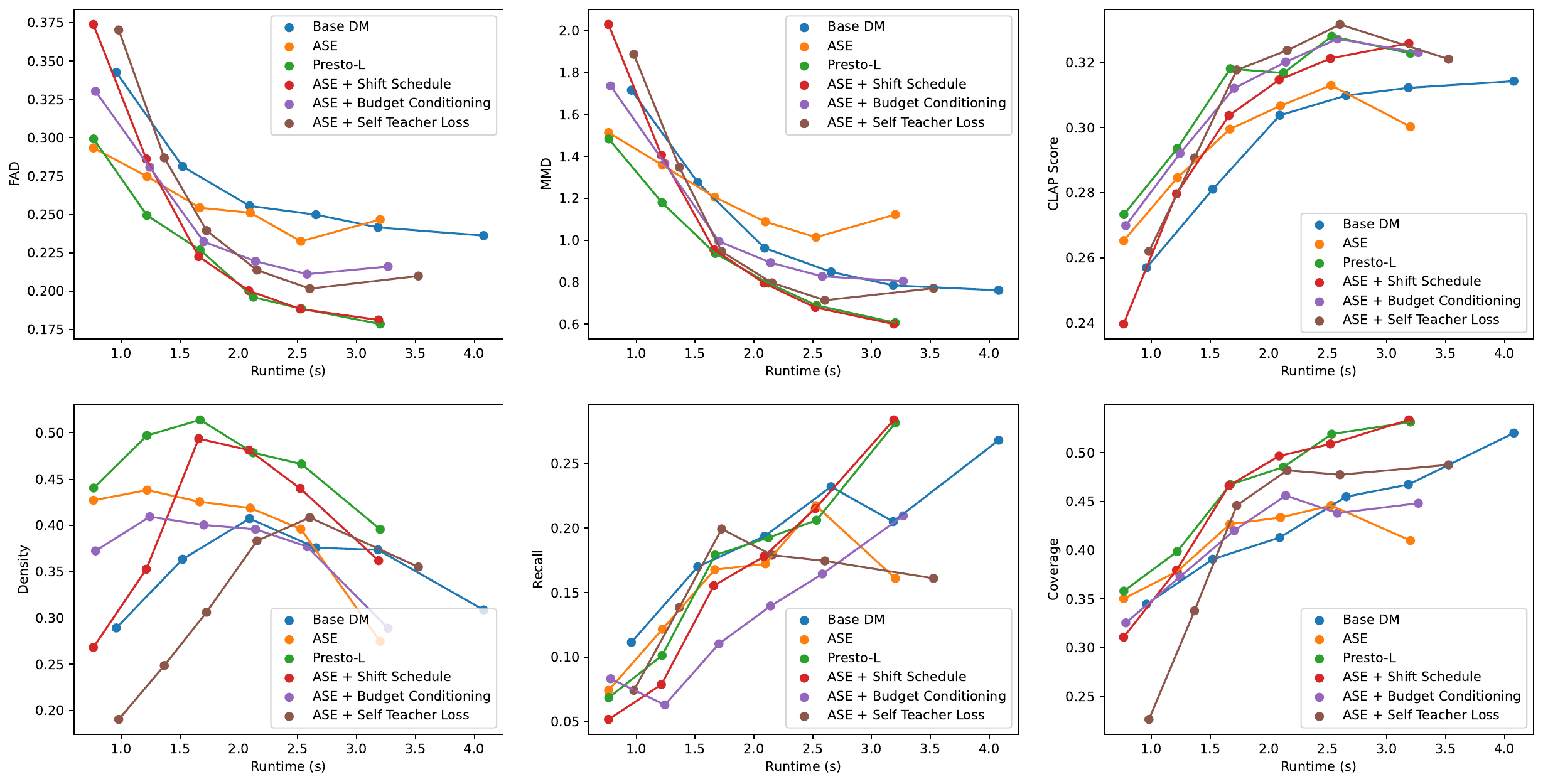}
    \caption{\textbf{Presto-L} ablation. Each individual change of our layer distillation vs ASE is beneficial.}
    \label{fig:l_abl}
\end{figure}
To investigate how each facet of our \textbf{Presto-L} method contributes to its strong performance vs. ASE, we ran an additional ablation combining ASE with each component (i.e.~the shifted dropping schedule, explicit budget conditioning, and the self-teacher loss). 
In \fref{fig:l_abl}, we see that the core of \textbf{Presto-L}'s improvements come from the shifted dropping schedule (which preserves final layer behavior), as the ASE+shift performs similarly to \textbf{Presto-L} on high-step FAD and MMD. Additionally, we find that the budget conditioning and self-teacher loss help text relevance more so than the shifted schedule does. All together, the combination of \textbf{Presto-L}'s design decisions leads to SOTA audio quality (FAD/MMD/Density) and text relevance compared to any one facet combined with ASE.

\subsection{Discrete-Time Failure Modes}
\begin{figure}[h!]
    \centering
    \includegraphics[width=\linewidth]{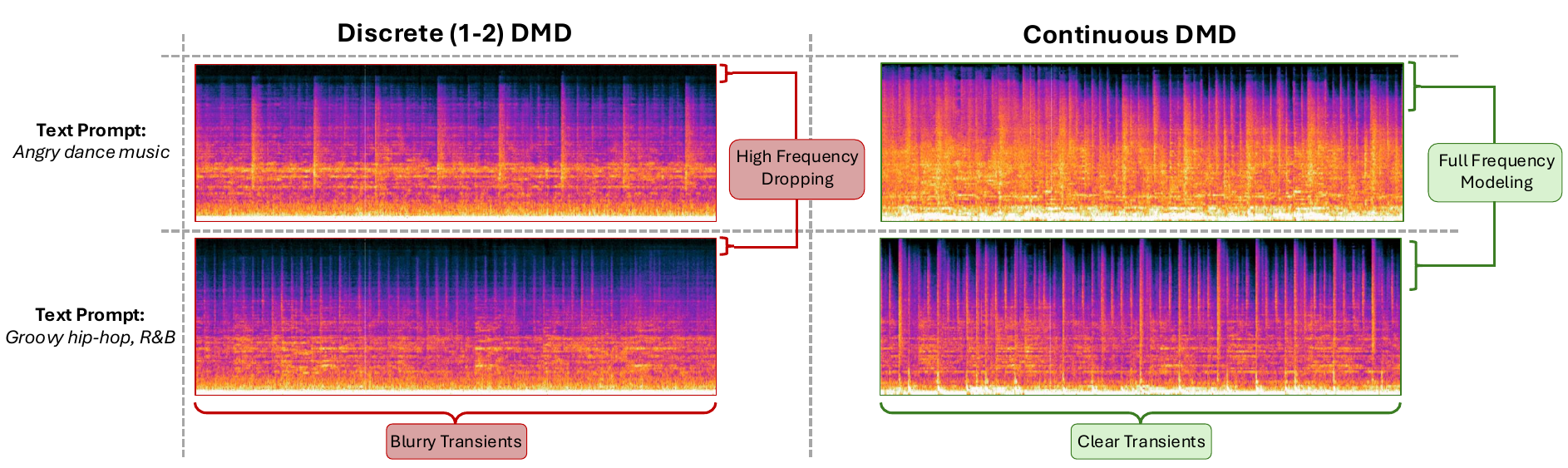}
    \vspace{-0.6cm}
    \caption{Failure mode of 1-2 step discrete models vs. continuous models (each row is same random seed and text prompt), with 2-step generation. Hip-Hop adjacent generations noticeably drop high frequency information, and render percussive transients (hi-hats, snare drums) poorly.}
    \label{fig:hifreq}
\end{figure}

In~\fref{fig:hifreq}, we visualize the poor performance of distilled models that use 1-2 step discrete-time conditioning signals. Notice that for the same random seed, the high-frequency performance is visually worse for discrete-time vs. continuous-time conditioning, motivating our proposed methods.

\subsection{Listening Test Results}
We visualize our listening test results from~\sref{sec:listening_test_main} using a violin plot. 

\begin{figure}[h!]
    \centering
    \vspace{-.1cm}
    \includegraphics[width=0.4\textwidth]{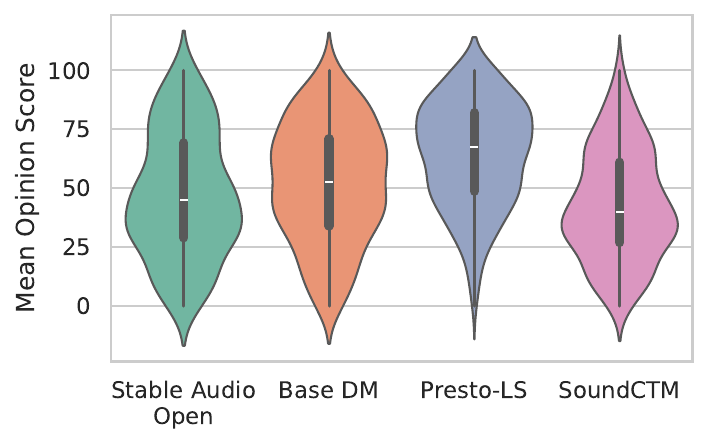}
    \vspace{-.25cm}
    \caption{Violin plot from our listening test. Presto-LS is preferred over other baselines ($p<0.05$).}
    \label{fig:mushra}
\end{figure}

\subsection{Rejection Sampling Results}
We show rejection sampling results where we generate a batch during inference and then use CLAP to reject the $r$ least similar generations to the input text prompt. CLAP rejection sampling  improves CLAP Score and maintains (and sometimes \emph{improves}) FAD and MMD, but reduces diversity.
\begin{figure}[h!]
    \centering
    \includegraphics[width=\textwidth]{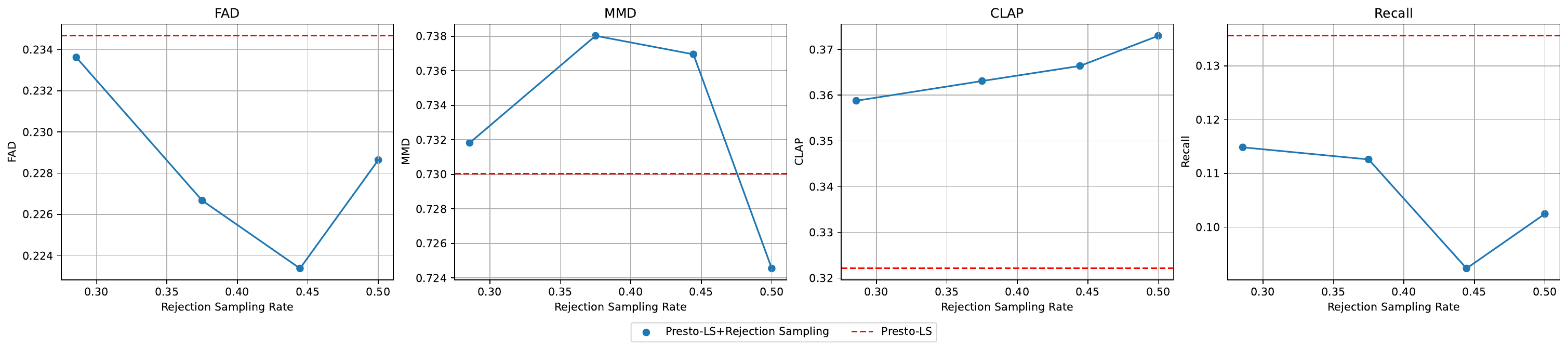}
    \vspace{-.8cm}
    \caption{Rejection sampling eval metrics vs. rejection ratio. Base \textbf{Presto-LS} in red. CLAP rejection sampling improves both CLAP score and overall quality, while reducing diversity.}
    \label{fig:rej}
\end{figure}

\end{document}

%% file: math_commands.tex
\usepackage{amsmath,amsfonts,bm}

\def\eqref#1{equation~\ref{#1}}

\def\1{\bm{1}}

\DeclareMathAlphabet{\mathsfit}{\encodingdefault}{\sfdefault}{m}{sl}
\SetMathAlphabet{\mathsfit}{bold}{\encodingdefault}{\sfdefault}{bx}{n}